\documentclass[twocolumn,showkeys,preprintnumbers,amsmath,amssymb,nofootinbib]{revtex4-1}

\usepackage{graphicx,paralist,comment}

\usepackage{hyperref}
\usepackage[hyphenbreaks]{breakurl}

\newcommand\sbullet[1][.75]{\mathbin{\vcenter{\hbox{\scalebox{#1}{$\bullet$}}}}}

\newcommand\apjl{Astrophys. J. Lett.}
\newcommand\aj{Astron. J.}

\newcommand\grl{Geophys. Res. Lett.}
\newcommand\physrep{Phys. Rep.}

\begin{document}

\title{Science opportunities with solar sailing smallsats}

\author{Slava G. Turyshev$^1$,
Darren Garber$^2$,
Louis D. Friedman$^3$,
Andreas M. Hein$^{4}$,
Nathan Barnes$^5$,
Konstantin Batygin$^6$,\\
G. David Brin$^7$, 
Michael E. Brown$^6$,
Leroy Cronin$^{8}$,
Artur R. Davoyan$^{9}$,
Amber Dubill$^{10}$,
T. Marshall Eubanks$^{11}$,\\
Sarah Gibson$^{12}$,
Donald M. Hassler$^{13}$,
Noam R. Izenberg$^{10}$,
Pierre Kervella$^{14}$,
Philip D. Mauskopf$^{15}$,
Neil Murphy$^1$,\\
Andrew Nutter$^{16}$,
Carolyn Porco$^{17}$,
Dario Riccobono$^{18}$,
James Schalkwyk$^{19}$,
Kevin B. Stevenson$^{10}$,
Mark V. Sykes$^{20}$,\\
Mahmooda Sultana$^{21}$,
Viktor T. Toth$^{22}$,
Marco Velli$^{23}$, and
S. Pete Worden$^{19}$
}

\affiliation{\vskip 5pt
$^1$Jet Propulsion Laboratory, California Institute of Technology,4800 Oak Grove Drive, Pasadena, CA 91109-0899, USA}

\affiliation{\vskip 3pt
$^2$NXTRAC Inc., Redondo Beach, CA 90277, USA}

\affiliation{\vskip 3pt
$^3$The Planetary Society (Emeritus), Pasadena, CA 91101, USA}

\affiliation{\vskip 3pt
$^{4}$SnT, University of Luxembourg, Luxembourg}

\affiliation{\vskip 3pt
$^5$L'Garde Inc., Tustin, CA 92780, USA}

\affiliation{\vskip 3pt
$^6$Division of Geological and Planetary Sciences, California Institute of Technology, Pasadena, CA 91125, USA}

\affiliation{\vskip 3pt
$^{7}$Futures Unlimited,  2240 Encinitas Blvd - D-300, Encinitas, CA 92024 USA}

\affiliation{\vskip 3pt
$^{8}$School of Chemistry, University of Glasgow, G12 8QQ, UK}

\affiliation{\vskip 3pt
$^{9}$Department of Mechanical and Aerospace Engineering, University of California, Los Angeles, USA}

\affiliation{\vskip 3pt
$^{10}$Johns Hopkins University Applied Physics Laboratory, \\
11100 Johns Hopkins Road, Laurel, Maryland 20723-6099, USA}

\affiliation{\vskip 3pt
$^{12}$Space Initiatives Inc, Newport, VA 24128, USA}

\affiliation{\vskip 3pt
$^{11}$UCAR, 3090 Center Green Dr., Boulder, CO 80301, USA}

\affiliation{\vskip 3pt
$^{13}$Southwest Research Institute, 1050 Walnut St., Boulder, CO 80302, USA}

\affiliation{\vskip 3pt
$^{14}$LESIA, Observatoire de Paris, Universit\'e PSL, CNRS, Sorbonne Universit\'e, Universit\'e Paris Cit\'e, 5 place Jules Janssen, 92195 Meudon, France}

\affiliation{\vskip 3pt
$^{15}$Department of Physics and School Of Earth and Space Exploration, Arizona State University, Tempe, AZ 85287, USA
}

\affiliation{\vskip 3pt
$^{16}$Gama Space, 128 bis Avenue Jean Jaur\'es, 94200 Ivry-sur-Seine, France
}

\affiliation{\vskip 3pt
$^{17}$Visiting Scholar, University of California, Berkeley, CA 94720, USA}

\affiliation{\vskip 3pt
$^{18}$Argotec, 52 Via Cervino, Turin, 10155, Italy}

\affiliation{\vskip 3pt
$^{19}$Breakthrough Initiatives, Building 18, Second Floor, PO Box 1, Moffett Field, CA 94035, USA}

\affiliation{\vskip 3pt
$^{20}$Planetary Science Institute,
1700 E. Fort Lowell, Suite 106
Tucson, AZ 85719, USA}

\affiliation{\vskip 3pt
$^{21}$NASA Goddard Space Flight Center, 8800 Greenbelt Rd., Greenbelt, MD 20771, USA}

\affiliation{\vskip 3pt
$^{22}$Ottawa, Ontario K1N 9H5, Canada}

\affiliation{\vskip 3pt
$^{23}$Department of Earth, Planetary and
Space Sciences, University of California, Los Angeles, USA}

\date{\today}

\begin{abstract}

Recently, we witnessed how the synergy of small satellite technology and solar sailing propulsion enables new missions. Together, small satellites with lightweight instruments and solar sails offer affordable access to deep regions of the solar system, also making it possible to realize hard-to-reach trajectories that are not constrained to the ecliptic plane. Combining these two technologies can drastically reduce travel times within the solar system, while delivering robust science. With solar sailing propulsion capable of reaching the velocities of $\sim$5--10\,AU/yr, missions using a rideshare launch may reach the Jovian system in two years, Saturn in three. The same technologies could allow reaching solar polar orbits in less than two years. Fast, cost-effective, and maneuverable sailcraft that may travel outside the ecliptic plane open new opportunities for affordable solar system exploration, with great promise for heliophysics, planetary science, and astrophysics. Such missions could be modularized to reach different destinations with different sets of instruments. Benefiting from this progress, we present the ``Sundiver'' concept, offering novel possibilities for the science community. We discuss some of the key technologies, the current design of the Sundiver sailcraft vehicle and innovative instruments, along with unique science opportunities that these technologies enable, especially as this exploration paradigm evolves. We formulate policy recommendations to allow national space agencies, industry, and other stakeholders to establish a strong scientific, programmatic, and commercial focus, enrich and deepen the space enterprise and broaden its advocacy base by including the Sundiver paradigm as a part of broader space exploration efforts.

\end{abstract}

\maketitle

~
\vskip 5em

\onecolumngrid

\begin{center}
\begin{minipage}{0.7\linewidth}
\tableofcontents

\vskip 1em

\end{minipage}
\end{center}
\twocolumngrid

\section{\label{sec:intro}Introduction}

The exploration of the outer solar system began on March 2, 1972, with the launch of the Pioneer 10\,\footnote{\url{https://en.wikipedia.org/wiki/Pioneer_10}} spacecraft on humanity's first mission to the planet Jupiter. Pioneer 10 became the first artificial object to achieve the escape velocity needed to leave the solar system, paving the way for other missions to deep space.

In the fifty years since, NASA has only five other spacecraft beyond the orbit of Jupiter (Pioneer 11\,\footnote{\url{https://en.wikipedia.org/wiki/Pioneer_11}}, Voyager 1/2\,\footnote{\url{https://voyager.jpl.nasa.gov/}}, Cassini--Huygens\footnote{\url{https://solarsystem.nasa.gov/missions/cassini/overview/}} and New Horizons\footnote{\url{http://pluto.jhuapl.edu/}}) and no new missions are planned for at least another decade. Each of these spacecraft spent many years in design and construction, and took nearly a decade of transit time to reach their intended destinations. Instrument development was frozen years prior to launch, often resulting in outdated technology being flown. Because of their high costs and success requirements, these missions were inevitably engineered with multiple levels of redundancy, increasing their cost further. They were designed, built, and operated by large engineering and science teams, resulting in each mission costing billions of dollars.

Evidently, solar system exploration requires significant time and money. Under the current paradigm, if we wish to visit Uranus again, we need to start planning decades in advance. Designing and building a craft to cross the billion-plus miles is likely to take over a decade, and the cruise to the solar system's third-largest planet could require an additional 15 years. Facing this reality, the National Academies of Sciences, Engineering and Medicine (NASEM) in its report ``{\em Origins, Worlds, and Life: A Decadal Strategy for Planetary Science and Astrobiology 2023-2032}'' \cite{NAP26522}, issued in April 2022, nevertheless recommended that the U.S. should launch a mission to Uranus in the late 2030s. It would be our first visit to the gas giant since 1986 and the first dedicated Uranus mission with an orbiter. Such a mission to the outer reaches of the solar system will likely cost more than \$4 billion.

In addition, an Enceladus orbiter/lander flagship mission was given the second highest priority for the upcoming decade by the solar system Decadal Survey committee, after a Uranus orbiter/probe mission. This is exciting because Enceladus has been targeted as one of the high priority ocean worlds for the search of life \cite{Hao-etal:2022}. That report became a blueprint for NASA, aiming to contribute to our understanding of the solar system.

However, there are two challenges: First, both the NASA budget and the plutonium needed would have to be increased to accommodate a second flagship later in the decade. And second, at the time of writing this paper, it is expected that the NASA budget will not be large enough to accommodate even the Uranus mission until later in the decade, delaying of course the arrival at Uranus by at least another decade.  These challenges will be difficult to resolve.

Beyond these two missions, there are many more exciting destinations and research objectives in the solar system and beyond, spanning many scientific disciplines, including planetary science, heliophysics and astrophysics, potentially leading to major discoveries in each of these areas. Although the science community had consistently argued in favor of such missions, NASA, given its current budget constraints, cannot afford them all. A substantial reason for the high costs is our reliance on slow and expensive chemical propulsion, operating at the limits of its capabilities, effectively rendering the current solar system exploration paradigm unsustainable. A new approach is needed.

Chemical propulsion puts an upper limit on the distances we can reach in a given time: $\sim$40~AU\footnote{The Sun Earth distance is defined as 1 Astronomical Unit (AU).} (Pluto's distance) within a decade; $\sim$500~AU (anticipated distance to the hypothetical {\em Planet 9}\,\footnote{\url{https://en.wikipedia.org/wiki/Planet_Nine}}) within a century. Alternative launch and propulsion methods, notably a heavy lift launch vehicle and nuclear propulsion, could achieve higher speeds and substantially reduce these durations, but both these technologies are very expensive developments, still at low-to-medium readiness levels, and already costing many billions of dollars.

Two new interplanetary technologies have advanced in the past decade to the point where they may enable inspiring and affordable missions to reach farther and faster, deep into the outer regions of our solar system: Interplanetary smallsats \cite{Staehle-etal:2012,Staehle-etal:2013,Norton-etal:2014}, the first of which have been demonstrated by JPL as MarCO\footnote{\url{https://www.jpl.nasa.gov/cubesat/missions/marco.php }} on the Mars InSight mission, and solar sails, which utilize solar radiation pressure for propulsion.

Recent major advances in solar sailing technology include the successful JAXA-built spacecraft IKAROS\footnote{\url{https://global.jaxa.jp/countdown/f17/overview/ikaros_e.html }}, which demonstrated interplanetary solar sail technology on a Venus-bound mission launched in 2010. In 2019, a successful orbital demonstration by the LightSail-2\,\footnote{\url{https://www.planetary.org/explore/projects/lightsail-solar-sailing/}} flight, led by The Planetary Society, raised confidence in solar sails and paved the way for the NASA interplanetary mission NEA-Scout\footnote{\url{https://www.nasa.gov/content/nea-scout/}} that was launched in 2022. Japan is now developing OKEANOS\footnote{\url{https://en.wikipedia.org/wiki/OKEANOS}} as a follow-up to IKAROS for outer planet missions. These efforts have improved sail materials, sail deployment mechanisms, propulsion thrust vector control, high-speed radiation-hardened computers, and modular spacecraft components, all driving down mass, risk, and cost.

Solar sails provide very large $\Delta v$ (potentially several ten km/s) enabling new vantage points for science observations that are inaccessible or impractical using conventional chemical or electric propulsion. Solar sails obtain thrust by using highly reflective, lightweight materials that reflect sunlight to propel a spacecraft while in space. The continuous photon pressure from the Sun provides thrust, eliminating the need for heavy, expendable propellants employed by conventional on-board chemical and electric propulsion systems, which limit mission lifetime and observation locations. Sails cost relatively little to implement, and the continuous solar photon pressure provides the thrust needed to perform a wide range of advanced maneuvers, such as hovering indefinitely at points in space (``stationkeeping in unstable orbits''), or conducting high $\Delta v$ orbital plane changes. Both of these capabilities are highly desirable for out-of-the ecliptic heliophysics (e.g., solar polar) and space weather monitoring missions (e.g., Sun-Earth line sentinels).

Solar sail propulsion systems can also accelerate a spacecraft to speeds that are much greater than what can be achieved using present-day propulsion systems. However, although they require no propellant, as the spacecraft acceleration is proportional to the sail area divided by the spacecraft mass, sails of practical size can only carry spacecraft of limited mass, i.e., smallsats.

Solar sails do not require a dedicated launch, which reduces mission cost. They can be part of a rideshare on a GEO or lunar mission, for example. By changing the attitude of the sail, the spacecraft can change orbit and move closer or farther away from the Sun. This is important especially for deep space outer solar system missions, since the optimal trajectory is to first get as close to the Sun as the materials and instruments allow, to harvest as much photonic momentum as possible as the vehicle accelerates away from the Sun. On the final approach to the Sun, the spacecraft would transition to a high-energy hyperbolic trajectory to the outer solar system and beyond. With today's technology, we can achieve speeds\footnote{Note that 1~AU/yr $\simeq$ 4.74 km/s.} of $\sim$7~AU/yr ($\sim$33~km/s) \cite{Turyshev-etal:2020-PhaseII}.

As part of the 2020 Phase III NASA Innovative Advanced Concepts (NIAC) study \cite{Turyshev-etal:2020-PhaseIII}, a Technology Demonstration Mission (TDM) of such an advanced sail was recommended to prove the enabling technology for reaching the solar gravitational lens (SGL) focal region, which begins at 547~AU from the Sun, in less than 25~years, requiring speeds greater than 20~AU/yr. This TDM will provide a demonstration of interplanetary smallsat--sailcraft capabilities. Once the basic TDM is accomplished, the concept can be adapted for use in different science missions. For example, steering the sail permits the change in inclination that is needed to achieve solar polar orbits, out-of-the-ecliptic missions, or rendezvous with interstellar objects (ISOs) \cite{Garber2022}. Missions could also target the outer solar system quickly, including flights to potentially life-supporting bodies, such as Enceladus, and beyond, to the neglected ice giants, Uranus and Neptune. This has led to the new paradigm for solar system exploration: The Sundiver concept, fast, cost-effective, frequent science missions with unconstrained trajectories.

After validation and with the right instruments, such missions can carry out investigations with broad science appeal, benefiting in particular the heliophysics, astrophysics and planetary science communities. In this paper, we address some of these opportunities.
In Section~\ref{sec:paradigm}, we present the SGL technology demonstration mission (TDM) that was developed under a NIAC Phase III effort \cite{Turyshev-etal:2020-PhaseIII} and became the prototype for the Sundiver mission concept.
In the sections that follow, we present some of the specific science opportunities offered by Sundivers:
heliophysics (Section~\ref{sec:helio}),
planetary science (Section~\ref{sec:planetary}),
and astrophysics, ultimately journeying to the focal region of the solar gravitational lens (Section~\ref{sec:astro}).
In Section ~\ref{sec:moon}, we briefly discuss some of the potential uses of the advanced solar sailcraft   in the ongoing lunar exploration efforts.
In Section~\ref{sec:concl}, we discuss the next steps in implementing the Sundiver mission concept and offer some relevant policy recommendations.
For convenience, we put some technically relevant material in the Appendices.
Appendix~\ref{app:techready} discusses current technology readiness of various components and subsystems;
Appendix~\ref{app:nearterm} presents some of the anticipated near-term developments.

\section{A new solar system mission paradigm}
\label{sec:paradigm}

\subsection{SGL technology demonstration mission (TDM)}
\label{sec:tdm}

\begin{figure*}
\includegraphics[scale=0.2588]{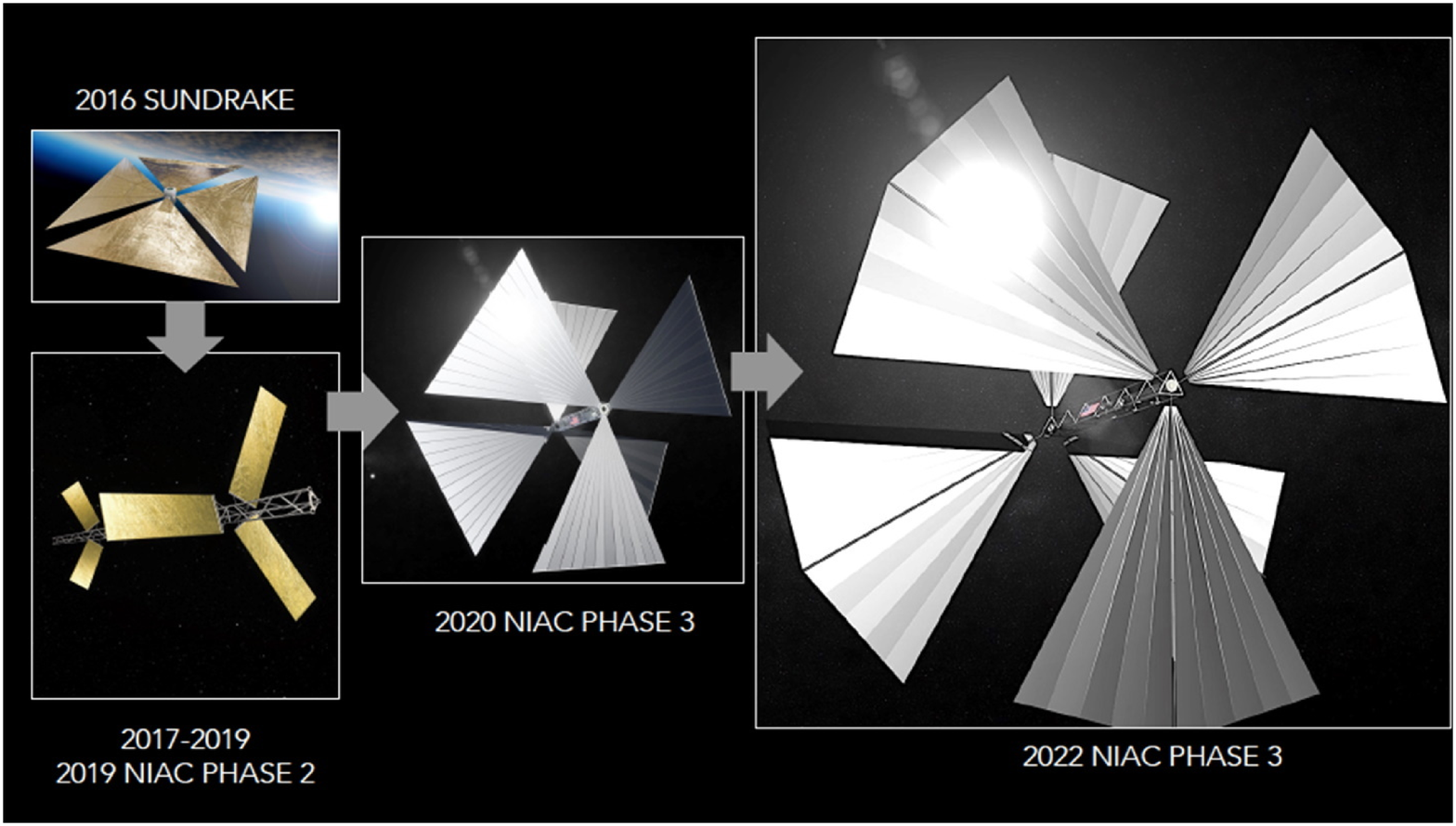}~\includegraphics[scale=0.24]{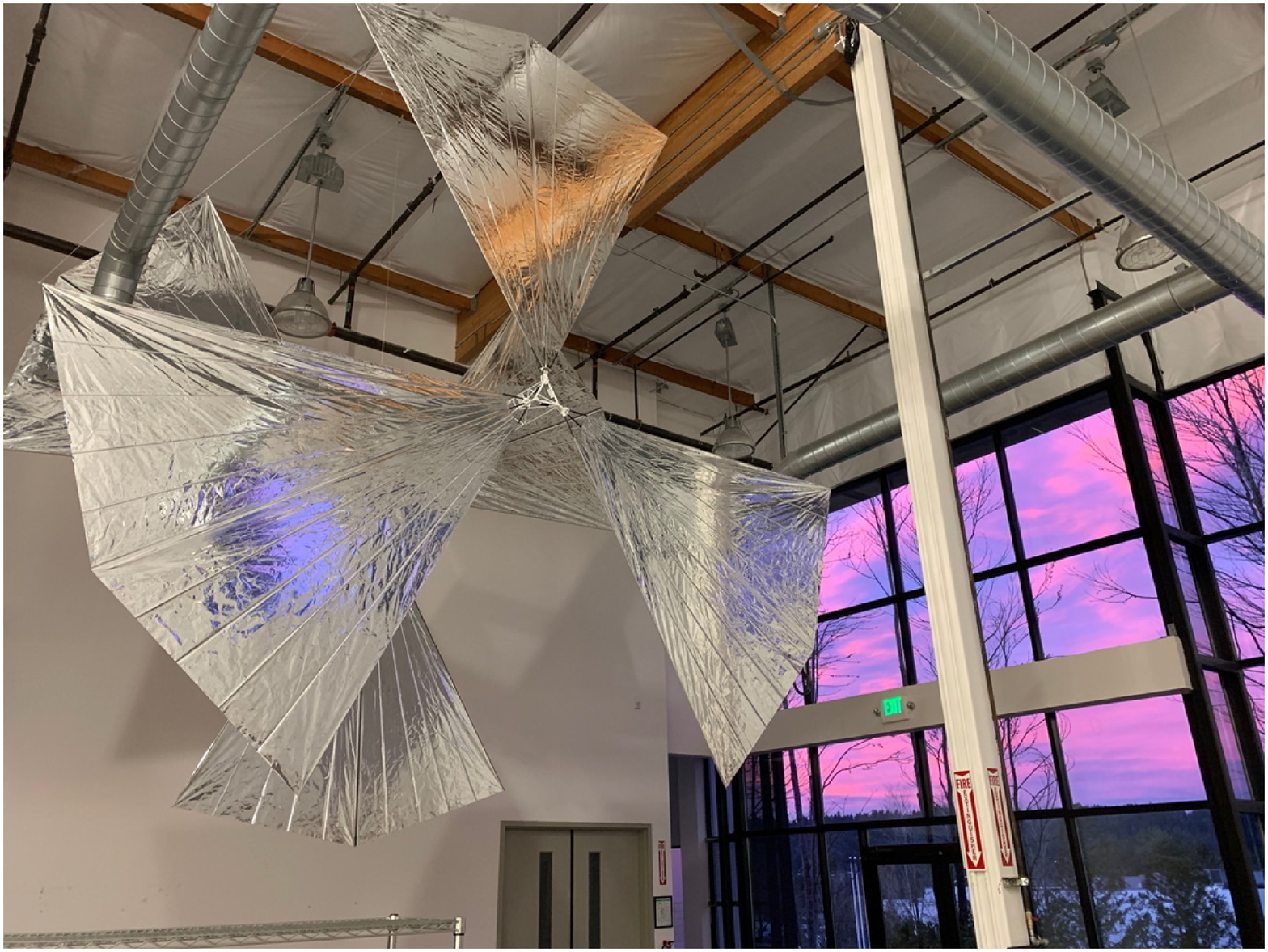}
\caption{\label{fig:TDM} Left: The LightCraft vehicle design evolution during the period of 2016-2022. Right: The 1:3 scale model of the TDM vehicle built at L'Garde, Inc. Tustin, CA, shown on display at the Xplore, Inc. facility in Redmond, WA, in March 2023.}
\end{figure*}

A Technology Demonstration Mission (TDM) was developed under a NIAC Phase III effort \cite{Turyshev-etal:2020-PhaseIII} as a first step to prove the feasibility of solar sailing interplanetary smallsats to fly through and out of the solar system at very high speeds to reach the solar gravity lens focal region at $>$600AU in less than 25 years and conduct direct multipixel imaging of exoplanets. The TDM vehicle is proposed as a series of TDM-preparatory flights in which incremental system builds will be designed, built, tested and flown, thus validating the technology and mission operations. The current design, LightCraft, evolved from the 2016 Sundrake vehicle. It is simpler to manufacture as a result of the lessons learned by building a $1:3$ model\footnote{The model is currently on display at Xplore's facility in Redmond WA, see \url{https://www.xplore.com/}.}, as shown in Fig.~\ref{fig:TDM} and Table~\ref{tb:TDMobj}. The TDM is the basis for the conceptual missions described in this paper (as discussed in Sec.~\ref{sec:sdconcept}).

The TDM is a 1--2~year test flight capable of carrying 1--2~kg of instrumentation. The solar system exit speed would be $>$5~AU/yr.  The LightCraft could include a set of simple experiments, such as optical communication transceivers, or a study of the behavior of sample sail materials in space. In the inner solar system, power is provided by photovoltaic elements placed on the sail. Beyond the specific experiments, the LightCraft demonstration flight will show a novel and cost-effective approach to explore the solar system by combining a low-cost launch, high sailcraft maneuverability, and rapid travel times.

The TDM sailcraft is a full 3-axis controlled interplanetary capable small spacecraft. Each sail element, or vane, can also be articulated to provide fine control to both the resultant thrust from solar radiation pressure and the vehicle's attitude. Each dynamic vane element is also a multifunctional structure hosting photovoltaics and communication elements with the requisite degrees of freedom to meet competing operational and mission requirements. The current TDM design total vane area is 120~m${}^2$ and the mass of the integrated TDM vehicle is 5.45~kg, resulting in an area-to-mass ratio of $A/m = 22$~m${}^2$/kg, or nearly 3 times the performance of other existing and planned sailcraft.

The TDM design is scalable to propel payloads up to 35--50~kg. Launching such payloads to high solar system transit velocities will require placing the vehicle on a trajectory with a close solar perihelion. Developments of new sail materials \cite{Davoyan2021} that can access  perihelion at 15--20\,$R_\odot$ is ongoing. These should be available already within this decade, enabling transit velocities 15--25~AU/yr, needed for missions to the deep regions of the solar system and ultimately to the SGL focal region.

The key design efforts for the TDM were focused on determining the requirements necessary to achieve the greater than 5~AU/yr hyperbolic egress velocity. This is substantially faster than Voyager in its interstellar flight and would set a speed record, demonstrating the concept for high solar system exit velocities. It will be sufficient to prove the viability of the new paradigm for fast solar system science missions that are being considered for the Sundiver sailcraft.

\begin{figure}
\includegraphics[width=\linewidth]{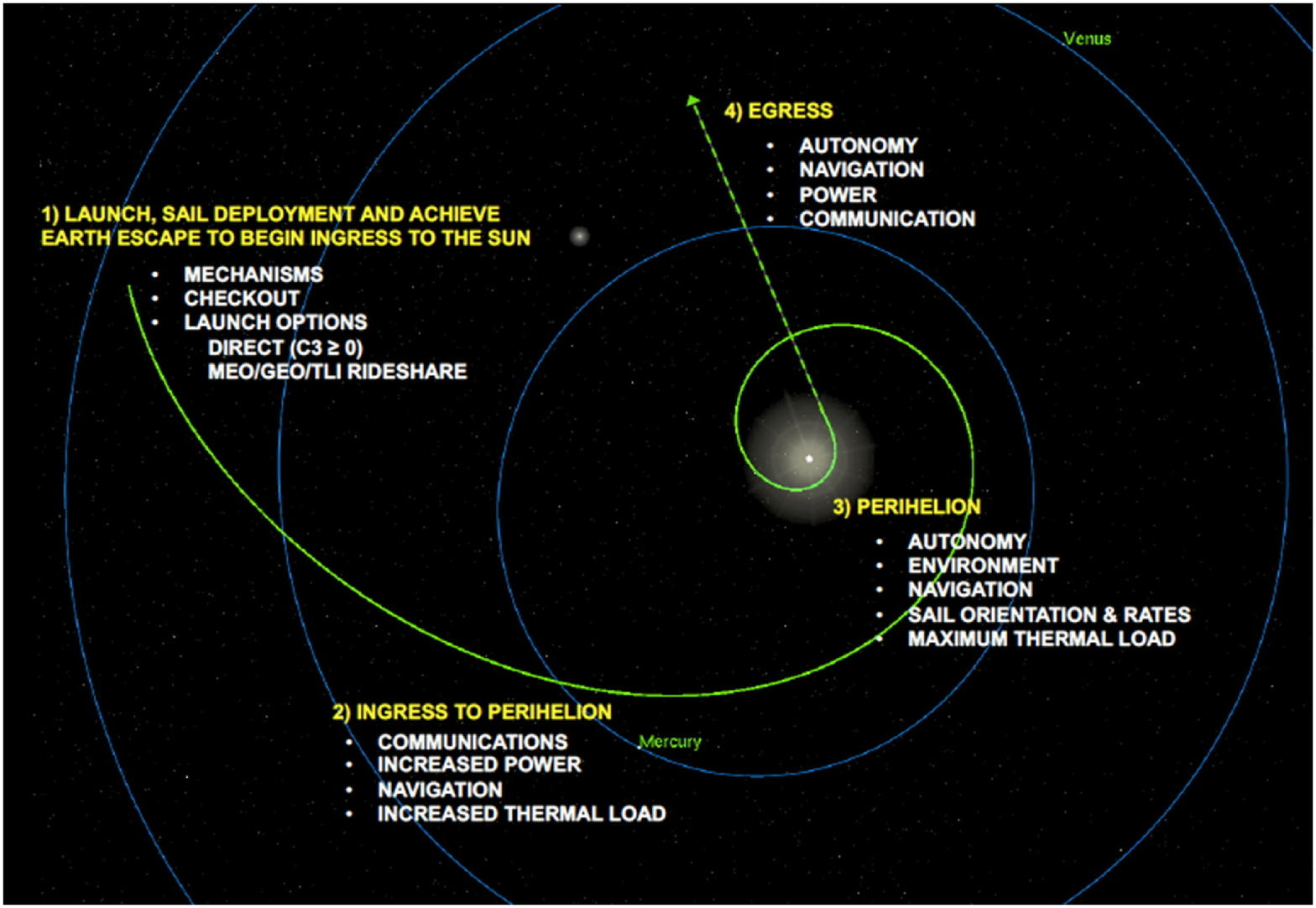}
\caption{\label{fig:SGLFTDM}Common TDM mission phases and systems engineering objectives. Trajectory plot shown is for the SGL mission.}
\end{figure}

The TDM design reference mission is an 18-month mission \cite{Garber2022b}, starting with an initial deployment from a GEO rideshare launch into a super-synchronous orbit and then, after checkout, an outspiral into interplanetary space, as shown in Fig.~\ref{fig:SGLFTDM}. Once Earth escape is achieved, the TDM vehicle will accelerate sunward to reach a perihelion at 0.24~AU, where it pivots to gain velocity from the significant radiation pressure and egress at over 5~AU/yr. From the mission analysis, an initial set of vehicle and system requirements were established, to enable the design of each subsystem to address control, communications, power, structure, and thermal load of the vehicle, along with the ground support infrastructure required to execute the mission successfully.

The mission design (Fig.~\ref{tb:TDMobj}) utilizes full gravity models for the entire solar system on the TDM vehicle according to the JPL DE430 planetary ephemerides. The trajectory is achieved with three simple control laws to maneuver the vehicle from geosynchronous orbit to perihelion and then egress:
\begin{inparaenum}[1)]
\item maximum acceleration: align vanes perpendicular to the Sun to increase velocity;
\item no acceleration: align vanes edge-on to the Sun; and
\item maximum deceleration: align vanes so that the resultant force is opposite to the heliocentric velocity vector, to decrease orbital kinetic energy.
\end{inparaenum}

The most dynamic phase for the TDM vehicle is its exit from the Earth's sphere of influence. During this phase the vehicle is alternating between acceleration and no acceleration every half period, also considering eclipse periods. Once in interplanetary space, the vehicle can simply decelerate inward towards the Sun. At perihelion, the control law changes to reorient the vanes to maximize acceleration along the velocity vector to achieve the necessary egress speed. The control laws provide full six degrees of freedom (DOF) control to account for errors, uncertainties, and the constrained capabilities of the vane actuators. The required position and attitude knowledge are well within the capabilities of existing terrestrial tracking systems and onboard inertial sensors (e.g., fine sun sensor, star trackers and accelerometers).

\begin{table}[h!]
\caption{\label{tb:TDMobj}TDM design objectives.
}
\begin{tabular}{|p{8.2cm}|}
\hline
TDM technical objectives:\\\hline
{\footnotesize$\bullet$}~$A/m$ ratio: $>50~{\rm m}^2/{\rm kg};$\\
{\footnotesize$\bullet$}~Achieve 6--8~AU/yr exit velocity; \\
{\footnotesize$\bullet$}~Survive perihelion of 0.2 AU; \\
{\footnotesize$\bullet$}~Low-cost \& manufacturable;\\
{\footnotesize$\bullet$}~Capabilities-based, no development; \\
{\footnotesize$\bullet$}~Rideshare compatible.\\\hline\hline
TDM design features (c. 2023):\\\hline
{\footnotesize$\bullet$}~$A/m$: 22.3~${\rm m}^2/{\rm kg}$ (i.e., 3$\times$NEA Scout);  current design is scalable to reach $>50~ {\rm m}^2/{\rm kg};$\\
{\footnotesize$\bullet$}~Six 20 m${}^2$ vanes (775 g per vane, 5~$\mu$m Kapton);\\
{\footnotesize$\bullet$}~Carbon fiber truss (120 g).\\\hline\hline
Avionics \& GNC leverages MARCO\footnote{\url{https://www.jpl.nasa.gov/missions/mars-cube-one-marco}}:\\\hline
{\footnotesize$\bullet$}~500 g for X-band SRD, 3 wheels, 2 star trackers, battery;\\
{\footnotesize$\bullet$}~100 g for shape memory motors.\\\hline\hline
Total mass:\\\hline
{\footnotesize$\bullet$}~ 5.37 kg; 86\% of mass is in vanes.
\\\hline
\end{tabular}
\end{table}

To operate the TDM vehicle while achieving the necessary orientation to maintain the prescribed mission trajectory, the maximum rotation rate for the vanes at perihelion is 0.5~$''/$sec (12$^\circ$ per day). A key element of the TDM design is the fact that it is a full three-axis stabilized vehicle, and the vanes are used primarily for propulsion, communications, and power, not attitude control. The attitude determination and control system consist of lightweight components developed for small satellites, including 3 reaction wheels, gyro, star tracker, sun sensors and accelerometers. The reaction wheels provide the precision attitude control and agility for slewing the vehicle in yaw, pitch, and roll.

\begin{figure}
\includegraphics[width=\linewidth]{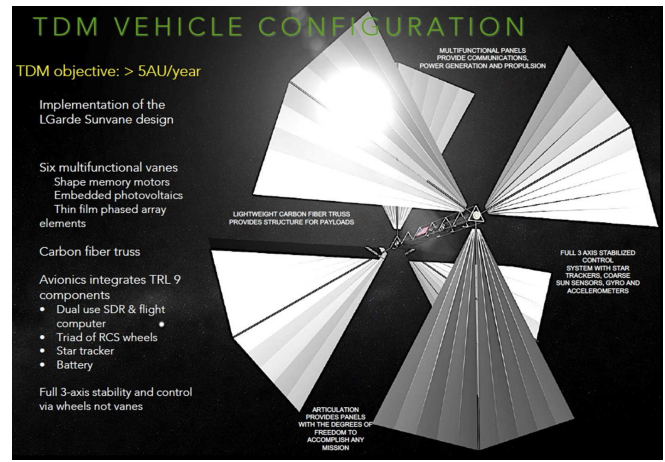}
\caption{\label{fig:TDMvehicle}TDM vehicle configurations (PDR: July 18, 2022) \cite{Garber2022b}.}
\end{figure}

The structure of the TDM vehicle (Fig.~\ref{fig:TDMvehicle}) consists of a 5~m carbon fiber truss hosting the guidance, navigation and control (GNC) components including a combined software-defined radio (SRD)  and flight computer, battery and to-be-determined payload. The current $A/m$ ratio for the TDM is 22~m$^2$/kg with further refinements in the design and corresponding schematics for the TDM vehicle, but also in the development of the workflow for vehicle manufacturing and testing\footnote{With the ongoing developmental efforts and existing sail materials this value was recently improved to 45--50 m$^2$/kg.}.

The current sail material on each vane is Kapton with an areal density of 3.55~g/m${}^2$. The current 20~m${}^2$ design requires 704~g of support structure resulting in 86\% of the TDM vehicle's mass being embodied within the vanes. Increasing the $A/m$ ratio provides a systems trade between increased acceleration versus maintaining the acceleration performance while increasing available payload mass. We identified other materials and variants of CP-1\,\footnote{CP-1 Polyamide is a high-performance material with various uses in display applications, space structures, thermal insulation, electrical insulators, industrial tapes, and advanced composites, \url{https://nexolve.com/advanced-materials/low-cure-polyimides/}} which have lower areal densities while enabling closer perihelion distances due to their exceptional thermal properties.

The thermal environment at 0.24~AU is close to the material limits for Kapton. All other orbit regimes for the TDM are manageable and easily accommodated. The use of the multi-layer insulation (MLI) and standard thermal treatments on the key surfaces of the vehicle led to a design that exceeded the thermal requirements for the sail, truss and bus, thus closing the design for the entire TDM vehicle.

Initial wiring and power sizing was completed for the current baseline components. The battery was sized to accommodate up to twice the potential eclipses in Earth orbit ($\sim$4 hours). The 10~W transmitter is the driving power component; however, a total of 120~m${}^2$ of total vane area is available for the placement of thin photovoltaic strips, which should be able to generate sufficient power to execute the mission. The EPS and avionics were recently modeled in a hardware in the loop simulator and tested with the GNC software following a defined mission trajectory (see Appendix~\ref{app:techready} for details on overall technological maturity).

Design trades investigated S, X and Ka-band communication link options for the TDM, including lightweight physical parabolic dishes and the use of thin film patch antennas as a phased array, utilizing the large surface area available on each vane. The current selection is based on incorporating phased array technology at X-band, given the availability of off-the-shelf components and their advantageous packaging and mass. Further investigations into potential Ka-band elements and varying vendors were conducted to validate the communication with the TDM vehicle up to 9~AU. Clearly, the same approach may be used to enable communication from much larger distances in the outer solar system.

These TDM development activities have demonstrated the feasibility of the vane structure to provide the necessary control authority during perihelion passage where the forces, torques, and environment are particularly harsh. While the TDM goal is to achieve a velocity of greater than 5~AU/yr, the threshold heliocentric exit velocity for the SGL focal mission is 20~AU/yr. This factor-of-four improvement in exit speed will be achieved by a combination of increasing the area-to-mass ratio by an equivalent factor of four and decreasing the perihelion distance by a factor of four.

The key technical challenge is to increase the payload capacity of the sailcraft from 1--2 kg to 10--15 kg and beyond, which may be achieved by in-flight assembly of an SGL spacecraft. The in-flight (as opposed to Earth-orbiting or cislunar) autonomous assembly \cite{Helvajian2022} allows us to build large spacecraft from modules, separately delivered in the form of microsats ($<$20 kg), where each microsat is placed on a fast solar system transit trajectory via solar sail propulsion to velocities of $\sim$10 AU/yr. Such a modular approach of combining various microsats into one larger spacecraft for a deep space mission is innovative and will be matured as part of the TDM flights. This unexplored concept overcomes the size and mass limits of typical solar sail missions. Autonomous docking and in-flight assembly are done after a large $\Delta v$ maneuver, i.e., after passing through perihelion. The concept also offers the compelling ability to assemble different types of instruments and components in a modular fashion, to accomplish many different mission types.

Some of the aspects of the in-flight assembly require functions that are not available among the mature docking technologies\footnote{Considering the recent NASA CPOD mission, rendezvous, proximity operations and docking technologies have achieved TRL~9 for microsats, see \url{https://www.nasa.gov/directorates/spacetech/small_spacecraft/cpod_project.html}.}. In particular, the docking mechanism shall guarantee structural, power and data connections between the modules (as performed by the docking mechanisms on the ISS). Another aspect that should be carefully considered for in-flight autonomous assembly is the high accuracy required for the egress trajectory determination and control. In fact, since each microsat is placed on an egress trajectory on its own and the assembly occurs afterwards, it is crucial to guarantee that all microsats are placed on the right egress trajectory.

The 2020 NIAC Phase III study concluded with a TDM Preliminary Design Review (PDR) on July 18, 2022 \cite{Garber2022b}. Next is pre-project mission development, which includes final design, hardware development, full-scale prototype construction, as well as hardware and software testing (see Tab.~\ref{tb:TDMobj} and Fig.~\ref{fig:TDMvehicle}). Should funding be available, the TDM Critical Design Review (CDR) may be conducted in November 2023, when flight project commitment is expected, including a firm costing of the TDM. The total project cost will depend on the selected mission objectives, science payload, and experiments, and is expected to be in the range of \$17--20M.

\subsection{The Sundiver concept}
\label{sec:sdconcept}

The TDM development is scalable to larger sail areas and hence higher payload capacities. It may be extended to enable other science missions in the solar system. There is a natural evolution of the TDM into the Sundiver concept: small and fast-moving sailcraft that could enable the exploration of distant regions of the solar system much sooner and faster than previously considered. Just like LightCraft, Sundiver vehicles will be solar-sail-driven smallsats that depart Earth, spiral in toward the Sun, fly through close perihelion and then outward from the Sun to desired destinations at speeds far exceeding that of any previous spacecraft. With the currently available sail materials, components, and instruments we can fly practical missions with speeds of up to 7~AU/yr, twice that of the current speed record holder, Voyager 1.

\begin{figure}
\includegraphics[width=\linewidth]{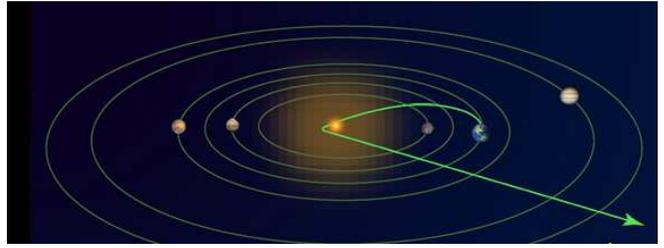}
\caption{\label{fig:sailtraj}Conceptual sailcraft trajectory.}
\end{figure}

\begin{figure*}
\includegraphics[width=0.75\linewidth]{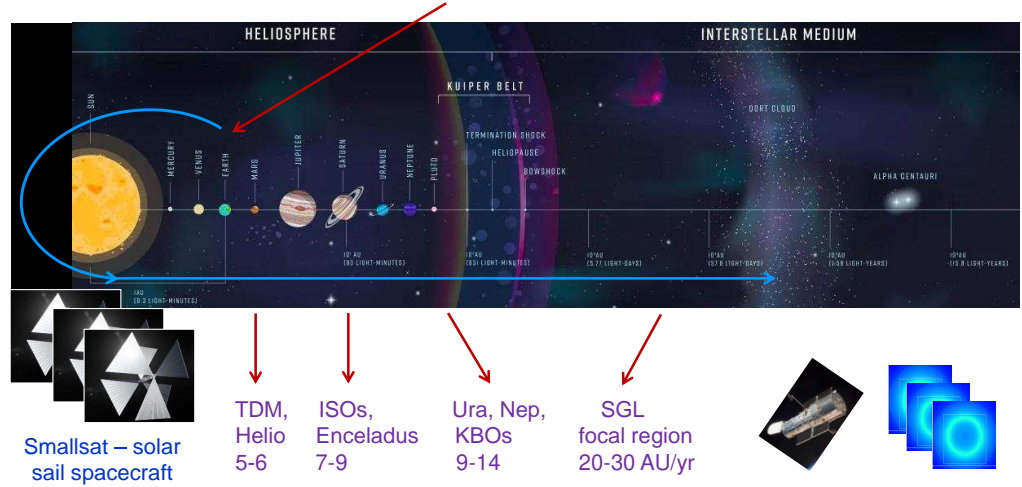}
\caption{\label{fig:TDMpara}New paradigm -- fast, low-cost, interplanetary sailcraft with trajectories unconstrained to the ecliptic plane. Note the capability development phases from TDM (at 5--6~AU/yr) to the mission to the focal region of the SGL (20--30~AU/yr).}
\end{figure*}

Technology requirements to address various science objectives can be assessed with respect to the technology readiness of the sail technology and that of other subsystems. The currently available sail and microsat bus technology could enable missions up to 5--7~AU/yr. Velocities larger than 7~AU/yr will require novel sail materials and technologies. With sail materials already being developed, smallsat velocities up to 20--25 AU/yr will be achievable in 5--7 years. The Sundiver concept therefore offers breakthrough capabilities that the science community has been  waiting for decades.

Missions in the inner solar system, $<$5~AU, may rely on the Sun to power their subsystems. For that, photovoltaic (PV) elements can be placed directly on the sail for reliable power generation. However, access to the outer solar system requires other power generation methods that do not require sunlight (e.g., lightweight small radioisotope power sources and batteries), which are currently in the development. Hence, missions going into deep space, $>$10~AU, require technology development and thus are on a different time horizon than missions destined to the inner solar system.

This new mission concept utilizing an interplanetary smallsat and a novel solar sail design, combined with a rideshare or small launch vehicle, all powered by small planar radioisotope power arrays, will enable low-cost missions to and through the solar system. A mission to either the Jupiter or Saturn systems makes use of passing by the Sun at a perihelion of 0.2~AU ($\sim$$40~R_\odot$) to achieve 47~km/s, reaching Jupiter in $\sim$20 months and Saturn in less than three years (see Fig.~\ref{fig:sailtraj}). A system study is needed for such a planetary science mission, but our preliminary estimate is that a payload of $\sim$5--10~kg can be accommodated on a smallsat.

The technology readiness assessment suggests a phased approach for increasingly more ambitious missions (Fig.~\ref{fig:TDMpara}). Based on currently available technologies (see Appendix~\ref{app:techready}) and anticipated near-term developments (see Appendix~\ref{app:nearterm}), Sundivers
may be implemented sequentially in three phases:
\begin{itemize}
\item {\bf\em Phase I} includes implementing missions to destinations in the inner solar system ($<$\,3~AU) and outside the ecliptic plane. The key enabling technologies here are solar sail materials, solar power via  PV elements embedded in the sail, RF communication, etc. -- all resulting in transit velocities of $\sim$5--7~AU/yr. Most of the critical technologies for these missions already exist at high Technology Readiness Level (TRL)\footnote{\url{https://www.nasa.gov/pdf/458490main_TRL_Definitions.pdf}}, allowing the implementation of these missions in the next 2--3 years at a cost of \$30--35M, depending on the target.

The onboard capabilities that will be implemented on Sundivers rely on a spacecraft bus that will use the sails as a multipurpose subsystem, including power generation (i.e., with the sail-embedded PV elements) and communication (i.e., relying on the in-flight formation of parabolic or flat phased-array antennas for RF links). These small systems may also rely on in-flight autonomous aggregation of science instruments from modules, separately delivered by a group of three-axis stabilized fully functional microsats ($<$20~kg each) which are placed on fast solar system transfer trajectories via solar sail propulsion to large velocities in excess of $\sim$10~AU/yr (see Appendix~\ref{app:techready} for details on the current TRL of various subsystems). Some relevant technologies for autonomous multi-agent space systems are already being developed (e.g., CARDE\footnote{NASA Cooperative Autonomous Distributed Robotic Explorers (CADRE) project:  \url{https://www.nasa.gov/directorates/spacetech/game_changing_development/projects/CADRE}}).

As noted, the hyperbolic velocity depends on sailcraft area to mass ratio and the temperature limits of the materials determining how close we can fly to the Sun. This relationship is shown in Figure \ref{fig:am-ratio}.

\begin{figure}[h]
\includegraphics[width=\linewidth]{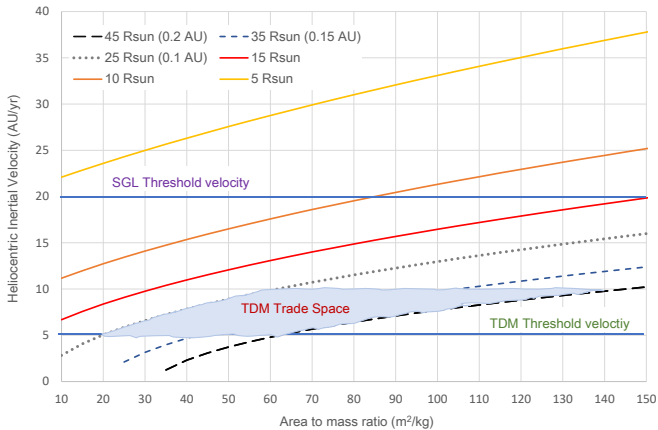}
\caption{\label{fig:am-ratio}Sailcraft exit velocity after solar
perihelion passage.}
\end{figure}

\item In {\bf\em Phase II}, technology developments already in progress will result in the maturation of several systems that will be required for deep solar system exploration. The technology roadmap includes advanced solar sail materials \cite{Davoyan2021}, larger sail deployment, on-board radioisotope power units, and hybrid RF/optical communications. Missions to distant solar system destinations at heliocentric distances of 5--40~AU with payloads of 15--25 kg can be flown in the next 5--8 years at a cost of \$50--85M. Missions to the more distant, medium heliocentric distances 100--200 AU (at least) will be enabled in this phase to provide for exciting science investigations \cite{Staehle-etal:2013,ISM2015,Staehle-etal:2020}.

As the solar sailing smallsats will be placed on very fast trajectories, placing Sundivers in orbit around a solar system body will be challenging. However they naturally yield several mission types including fast flybys, impactors, formation flights, and swarms. As the weight of the system is constrained, any instruments on board need to be small, lightweight, and low-power. Given the ongoing efforts in miniaturization of many instruments and subsystems, these challenges will be met by our industry partners who are already engaged in related technology developments.

\item These activities will be further extended in {\bf\em Phase~III}  with technologies for aggressive perihelion passages, planar radioisotope power units, and optical communications. Missions in this phase will be limited only by the available onboard power and will be able to reach destinations outside the solar system, in the interstellar medium, ultimately reaching the focal region of the SGL \cite{Helvajian2022}. For these missions to be able to operate beyond 200~AU, the development of small radioisotope power sources capable of powering communication and electric microthrusters for maneuvering in the SGL focal region will be critically important.

These sailcraft may be able to accommodate payloads over 25 kg, will have lightweight radioisotope elements for on-board power (see  \cite{Helvajian2022}), and will rely on optical communications. Depending on a particular science objective, such missions may cost \$90--100M, still competitive with the current paradigm.
\end{itemize}

Lightsail technology, combined with rideshare and deployment at Earth orbit, and the use of smallsat architecture components allow dramatic reduction in mission costs and lead time. LightSail-2, a \$7M low Earth orbit (LEO) solar sail mission launched in 2018, has operated successfully for two years in orbit. Also consider Solar Cruiser, a \$65M NASA interplanetary science mission that was originally scheduled to be launched in 2025 (as of 2022, the development is on hold for reasons unrelated to lightsail technology development). The relatively low cost of Solar Cruiser and LightSail-2 verify low-budget opportunities and short lead times (5--7 years). An assessment of the LightCraft 1 budget puts it in the same range. The current LightCraft 1 cost estimate, independently verified by Aerospace Corporation, is \$11M for its engineering model with a one-year interplanetary flight.

Unlike typical NASA programs where each mission is performed by a dedicated spacecraft, the LightCraft architecture can be repurposed to other missions and destinations, offering a novel model for space exploration. Many low-cost missions will be sent to diverse targets throughout the solar system. For example, the Jovian moons can be visited by multiple probes to ensure high throughput and validity of science data. Notably, with the use of the economies of scale, each successive mission will cost less than its predecessor.

Sundiver missions may also benefit from innovative solar sail technologies. One example is diffractive sails \cite{Swartzlander:2018,Swartzlander:2017, Swartzlander:2022,Swartzlander-etal:2022}, which may offer a potentially cost-effective path, especially for missions outside the plane of the ecliptic, such as solar polar orbiters \cite{Dubill-Swartzlander:2021}. Even the addition of smaller diffractive elements into a reflective sail can give major benefits to attitude control of a traditional sail \cite{Dubill:2020}. Other possibilities also include combining solar sails with the Oberth maneuver as close solar perihelia \cite{Bailer-Jones:2021}.

There are several science opportunities that may be realized with the Sundiver capabilities that are currently available or will become available by 2030. In the remainder of this paper, we discuss several candidate Sundiver solar system missions summarized in Table~\ref{tb:sci}, for convenience. The list is not exhaustive: there are likely many more science opportunities no one has yet considered. Nonetheless, the selection demonstrates the breadth of opportunities enabled by Sundivers.

\begin{table*}[!hbt]
\caption{\label{tb:sci} Candidate Sundiver missions that may be flown by 2030-2035.
Missions are presented according to their anticipated timelines within the Sundiver program,  their overall technology/concept readiness, and their preferred trajectories. For the solar system hyperbolic escape trajectories (HET), preferred spacecraft transit velocity, $v_{\tt tr}$, and desirable operational range, $r_{\tt ops}$, are  provided. Key parameters on some low thrust trajectories (LTT) are also shown. A mission concept readiness level (CRL) approach was used to indicate  overall feasibility/maturity of a concept.  CRLs are based on TRLs for systems, instruments, and mission architecture, and  range as CRL = ``low, medium, high, flight ready''. Note that, in addition to the solar polar imager mission, all astrophysics objectives will benefit from the capabilities of forming trajectories outside the ecliptic plane.}
\begin{tabular}{|p{6.8cm}|c|c|c|p{6.3cm}|}\hline
{\bf Science objective}&{\bf Section} & {\bf Timeline}&{\bf CRL}&{\bf Envisaged trajectory }\\\hline
\multicolumn{5}{l}{Heliophysics:}\\\hline
Studying the Sun with a solar polar imager\footnote{Several sailcraft for continuing monitoring are desirable.}& \ref{sec:spi} & Phase I &  high &  Solar polar orbit at $i\simeq 90^\circ$,  $r_{\tt ops}=0.4$ AU\\
Studying the heliosphere &  \ref{sec:heliosph} & Phase II & medium & HET:  $v_{\tt tr}\sim$ 7--10~AU/yr,  $r_{\tt ops}\simeq$ 100--150~AU \\
Probing the interstellar ribbon& \ref{sec:ribbon} &Phase II   &medium & HET:  $v_{\tt tr}>$ 12~AU/yr,  $r_{\tt ops}\simeq$ 100--250~AU\\
Studying the pristine interstellar medium & \ref{sec:pism} & Phase II-III & low & HET:
$v_{\tt tr}\sim$ 10--12~AU/yr,  $r_{\tt ops}>350$~AU \\
\hline
\multicolumn{5}{l}{Planetary science:}\\\hline
Study of the hard-to-reach asteroids\footnote{Precision trajectory pointing is needed.} & \ref{sec:astroids} & Phase I-II & high & LTT:  $v_{\tt tr}\sim$ 1--2~AU/yr,  $r_{\tt ops}\simeq$ 0.5--2.0~AU\\
Probing the plumes on Enceladus${}^{b,}$\footnote{At target, a slow flyby/orbiter with $v_{\tt tr}<$1~AU/yr is desirable.}  &  \ref{sec:plumes} & Phase II    & medium & HET:  $v_{\tt tr}>$ 3--5~AU/yr, $r_{\tt ops}\simeq9$~AU\\
Molecular biosignatures in the solar system & \ref{sec:mol-bio} &Phase II  & medium  &
HET: $v_{\tt tr}\sim$ 5--7~AU/yr,  $r_{\tt ops}\simeq $ 2--10~AU\\
Fast flybys of Uranus and Neptune &  \ref{sec:UrNep}& Phase II  & medium & HET: $v_{\tt tr}\sim$ 10--12~AU/yr,  $r_{\tt ops}\simeq$ 20--30~AU\\
Kuiper Belt and Oort Cloud objects & \ref{sec:kbooco} & Phase II-III & low &
HET: $v_{\tt tr}\sim$ 15--17~AU/yr,  $r_{\tt ops}\simeq$  90--250~AU \\
Probing Planet 9 & \ref{sec:planet9}& Phase III  & low & HET: $v_{\tt tr}>$ 17~AU/yr,  $r_{\tt ops}\simeq 380$~AU\\\hline
\multicolumn{5}{l}{Astrophysics:}\\\hline
Observing Earth as an exoplanet & \ref{sec:earth-exo}& Phase I & high & LTT: beyond Sun-Earth L2 \\
Intercepting and probing interstellar objects & \ref{sec:ico} & Phase II & medium & HET: $v_{\tt tr}\sim$ 5--7~AU/yr,  $r_{\tt ops}\simeq $ 5--12~AU\\
Zodiacal background and interplanetary dust\footnote{Trajectories outside ecliptic plane are preferred.} & \ref{sec:zodi} & Phase II  & medium & HET: $v_{\tt tr}\sim$ 7--10~AU/yr,  $r_{\tt ops}\simeq $ 5--20~AU\\
Cosmic background and the reionization epoch &  \ref{sec:cb-reionize}& Phase II & medium& HET: $v_{\tt tr}\sim$ 10--12~AU/yr,  $r_{\tt ops}\simeq $ 15--40~AU\\
Testing long range relativistic gravity
& \ref{sec:test_gravity} & Phase II & medium & HET: $v_{\tt tr}\sim$ 12--17~AU/yr,  $r_{\tt ops}\simeq $ 50--120~AU\\
Exoplanet imaging: the solar gravitational lens & \ref{sec:sgl} & Phase III & low & HET: $v_{\tt tr}\sim$ 23--25~AU/yr,  $r_{\tt ops}\simeq $ 650--900~AU\\\hline
\multicolumn{5}{l}{Cislunar and beyond:}\\\hline
Infrastructure development in cislunar space
 & \ref{sec:moon}  & Phase I & high & LTT in cislunar space; pole-sitters, etc.  \\\hline
\end{tabular}
\end{table*}

\section{Heliophysics}
\label{sec:helio}

Solar sails enable missions to observe the solar environment from unique vantage points, such as sustained observations away from the Sun-Earth line (SEL), which is of interest to a broad user community \cite{Gibson2018}; sustained sub-L1 (sunward of L1 along the SEL) stationkeeping for improved space weather monitoring, prediction, and science \cite{Denig2014} and supporting human spaceflight crew safety and health needs \cite{NAP11760}; sustained {\em in situ} Earth magnetotail measurements \cite{MacDonald2007}; and, in the midterm, observations that require a high-inclination solar orbit \cite{Kobayashi2020,Liewer2008}; Earth polar-sitting and polar-viewing observatories \cite{Ceriotti2011}; as well as multiple fast transit missions to study the transition between the heliosphere and the interstellar medium.

\subsection{Solar Polar Imager}
\label{sec:spi}

\begin{figure}
\includegraphics[width=0.9\linewidth]{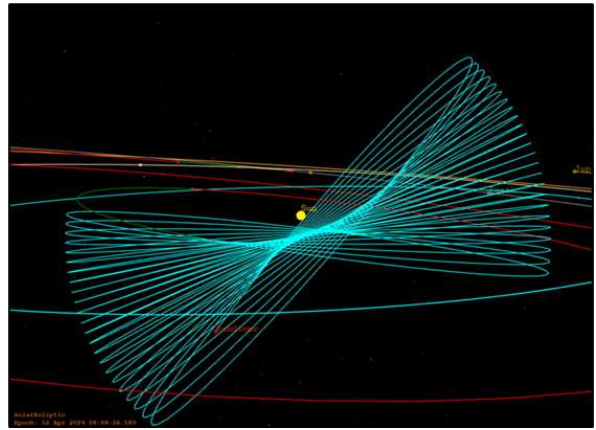}
\caption{\label{fig:SPItraj}Conceptual solar polar  trajectory. The ultimate science orbit is a 0.4~AU nearly circular heliocentric orbit with a heliographic inclination of 90$^\circ$.}
\end{figure}

Observations by past spacecraft such as Yohkoh\footnote{\url{https://www.isas.jaxa.jp/en/missions/spacecraft/past/yohkoh.html}}, ACE\footnote{\url{https://solarsystem.nasa.gov/missions/ace}}, Ulysses\footnote{\url{https://www.esa.int/Science_Exploration/Space_Science/Ulysses_overview}}, TRACE\footnote{\url{https://science.nasa.gov/missions/trace}}, SOHO\footnote{\url{https://soho.nascom.nasa.gov}} and the Parker Solar Probe\footnote{\url{http://parkersolarprobe.jhuapl.edu}}, and by current spacecraft such as RHESSI\footnote{\url{https://hesperia.gsfc.nasa.gov/rhessi3}}, Hinode\footnote{\url{https://www.nasa.gov/mission_pages/hinode}}, STEREO\footnote{\url{https://www.nasa.gov/mission_pages/stereo/main}}, Solar Orbiter\footnote{\url{https://www.nasa.gov/content/solar-orbiter-overview}}, and the Solar Dynamics Observatory (SDO)\footnote{\url{https://sdo.gsfc.nasa.gov/mission}} revolutionized our understanding of the Sun, its corona and the solar wind. Yet only one of these missions, Ulysses, had a high inclination, out-of-the-ecliptic orbit, and we have yet to send another mission to the solar poles, due to the high $\Delta v$ required for such out-of-plane maneuvers. As we have learned more about the Sun from these missions and the complement of ground-based telescopes, the need for more information from a polar perspective has only increased, as was demonstrated in the recent NASA Solaris Midscale Explorer Phase A Study \cite{Hassler:2020}. The Solar Polar Imager (SPI) mission concept \cite{Denig2014} uses a solar sail to place a spacecraft in a 0.48 AU circular orbit around the Sun with an inclination above 75$^\circ$, enabling high-latitude studies and extended direct observation of the solar poles (Fig.~\ref{fig:SPItraj}). Recently, a TDM-derived sailcraft with articulable vanes was used \cite{Garber2022b} to develop a mission concept to place a science spacecraft at 0.4~AU in a polar orbit around the Sun in less than 2 years.

{\bf\em Science objectives}: Observing the polar regions of the Sun with a combination of a Doppler magnetograph and coronal imagers yields opportunities for significant new science. Local helioseismology measurements of polar supergranulation flows, differential rotation, and meridional circulation, in addition to magnetograms, would allow us to understand the mechanisms of polar field reversals and the factors determining the amount of magnetic flux accumulating at the polar regions, which is a primary precursor of future sunspot cycles. Correlation between measurements of the Doppler signals in the polar regions with disk center measurements from the ground or from near-Earth spacecraft, such as SDO, could enable the determination of flows deep within the Sun. When Doppler and magnetograph observations are coupled to total solar irradiance monitoring, UV spectroscopic observations and {\em in situ} particle and field measurements, our knowledge of solar variability would be substantially enhanced. While Solar Orbiter will provide a glimpse of the polar regions\footnote{Solar Orbiter will reach an inclination of its orbit in the range of 24-33${}^\circ$, see \url{https://en.wikipedia.org/wiki/Solar_Orbiter}.}, it does not reach high enough solar latitudes to achieve the major scientific objectives defined for a potential solar polar mission such as Solaris \cite{Hassler:2020,Hassler:2021}.

Unique remote sensing and {\em in situ} observations made possible by an orbit reaching solar latitudes greater than 75$^\circ$ (Fig.~\ref{fig:SPItraj}) include:
\begin{inparaenum}[1)]
\item measurements of time-varying flows, including convection, differential rotation and meridional circulation in the polar regions of the Sun;
\item measurements of the polar magnetic field and its temporal evolution;
\item monitoring of Earth-directed coronal mass ejections from high latitudes;
\item observations of active regions over a significant fraction of their lifetimes;
\item measurements of the variation in the total solar irradiance with latitude;
\item measurements of chromospheric and low coronal outflow velocities as a function of structure and latitude; and
\item measurements of the variation in the magnetic fields, solar wind, and solar energetic particles (SEPs) with latitude at constant distance from the Sun.
\end{inparaenum}

{\bf\em Mission design and requirements}:
Even with the launch of the solar-polar focused Solaris mission in the near term, there remains a need  for sustained observations of solar physics processes on solar cycle timescales and, increasingly urgently, for monitoring and forecasting space weather. Both of these need continuous measurements from all latitudes and longitudes -- that is, $4\pi$ coverage of the Sun. The small solar sailcraft concept lends itself to a disaggregated approach wherein multiple launches could fill in this $4\pi$ around the Sun.

Because of weight restrictions, achieving all of the science objectives described in the previous section would require a swarm of smallsats. The most critical instruments for which to obtain full coverage are the Doppler magnetographs \cite{Hassler:2022}, which survey the Sun's surface magnetic field and surface and sub-surface flows, key to understanding how the magnetic field drives the solar cycle, the solar wind, and its frequent eruptions. A mission therefore would prioritize coverage for this instrument with at least four craft in orbit so that both poles are continuously observed. Other instruments, as described below, could be then launched to fill out the polar orbit like strings of pearls, providing observations to improve our understanding of the mechanism of solar activity cycles, polar magnetic field reversals, and the internal structure and dynamics of the Sun and its atmosphere.

The sailcraft would start from high Earth orbit and will continue towards the Sun using solar sailing. (To expedite the time to reach the initial perihelion, the mission may use additional propulsion.) After reaching the heliocentric distance of 0.25--0.3~AU, the sailcraft will stay at that distance and, by relying on its articulable vane design, will initiate the inclination change. It will proceed to raise the inclination by 2--3$^\circ$ every 21 days, yielding a polar orbit in two years or less. The ultimate orbit will be a polar 90$^\circ$ orbit at 0.4~AU from the Sun.

{\bf\em Instrumentation}:
Different science instruments include both {\em in situ} instruments (to measure the environment immediately surrounding the spacecraft, such as solar wind plasma---the electrified gas streaming from the Sun---and the electric and magnetic fields embedded within it) and remote sensing (to image the Sun).
{\em In situ} sensing instruments\footnote{Some instruments are discussed for the Solar Orbiter: \url{https://www.nasa.gov/content/solar-orbiter-instruments}} that may be accommodated on the Sundiver smallsats\footnote{See some of the relevant instruments are listed in \cite{Staehle-etal:2013}.} include:
\begin{inparaenum}[1)]
\item particle detector to measure energetic particles from the Sun over a wide range of energies with high temporal, energetic and mass resolutions;
\item a magnetometer to measure the strength and direction of the magnetic field around the spacecraft;
\item a solar wind analyzer to measure the charged particles that come from the Sun towards the spacecraft -- specifically, the electrons, protons and heavier particles that make up the bulk of the solar wind; and
\item instruments to measure the changes in the electric and magnetic fields around the spacecraft.
\end{inparaenum}

The mission could also accommodate several types of high-heritage remote sensing instruments, including
\begin{inparaenum}[1)]
\item Doppler magnetographs to survey the Sun's surface magnetic field and surface and sub-surface flows, key to understanding how the magnetic field drives the solar cycle, the solar wind, and its frequent eruptions  and already mentioned above as the primary science payload;
\item extreme UV/soft X-ray imagers to image solar coronal dynamic evolution;
\item spectroscopic/spectropolarimetric instruments capable of diagnosing plasma and magnetic properties from solar surface through lower atmosphere in visible and UV wavelengths;
\item white-light coronagraphs and heliospheric imagers to measure sunlight reflected off of solar wind electrons and track the structure and dynamics of the solar eruptions;
\item a hard X-ray spectrometer telescope to observe solar flare processes including the acceleration of particle to relativistic speeds, yielding crucial information about space weather as well as the underlying physics of how flares work.
\end{inparaenum}
With a disaggregated approach of multiple spacecraft bearing a variety of payloads, such instruments (remote-sensing and {\em in situ}) could work together to provide a comprehensive view of our star from yet unexplored viewpoints.

{\bf\em Technology readiness}:
The overall technological readiness for a
Phase I Heliophysics Sundiver smallsat is very high\footnote{In this respect, we anticipate that the  recommendations from the Decadal Survey for Solar and Space Physics (Heliophysics) 2024-2033 that is currently conducted by the NASEM, \url{https://www.nationalacademies.org/our-work/decadal-survey-for-solar-and-space-physics-heliophysics-2024-2033}, will include solar probe concepts that would greatly benefit from the solar sailing technology discussed here, especially because of its ability to form trajectories unconstrained to the ecliptic plane.}. In particular, a magnetometer could be sent to a point along the Sun-Earth line to demonstrate capabilities including hovering and stationkeeping as well as telemetry and communications sufficiency, solar power via embedded photovoltaics, and provide immediately useful space weather monitoring data, ``upstream'' of the Earth.  For an initial solar polar orbit smallsat, a small EUV imager ($\sim$3 kg) might be accommodated on a mission within the Phase I of the Sundiver program -- such an imager would enable images of the solar poles and demonstrate capabilities, such as pointing stability.

Once these initial demonstrations are made, and coupled with ongoing miniaturization of solar imaging and solar wind instrumentation, the capabilities for building a string of sensors distributed around the Sun (in $4\pi$ coverage) will be clear.

\subsection{Studying the heliosphere}
\label{sec:heliosph}

\begin{figure}
\includegraphics[width=0.8\linewidth]{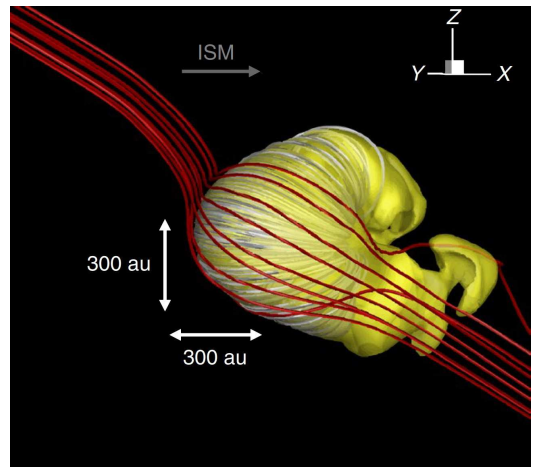}
\caption{\label{fig:heliosph}Two-lobe structure of the heliosphere. The white lines represent the solar magnetic field. The red lines represent the interstellar magnetic field. From \cite{Opher2020}.}
\end{figure}

As the Sun travels through the interstellar medium (ISM) it ejects plasma at speeds of 400--800~km/s. This solar wind flows well beyond the orbits of the planets and collides with the ISM. The bubble-like plasma region created by the solar wind around the Sun is called the heliosphere, and shields the solar system from cosmic ionizing radiation. At the boundary of the heliosphere, interaction of the solar wind with the interstellar gas creates an interface with a complex structure (Fig.~\ref{fig:heliosph}).

For decades, it was assumed that the heliosphere, with the Sun's motion relative to the ISM forming a long magnetotail, would have a shape similar to that of the Earth's magnetosphere, which resembles a comet's coma and tail, with a compressed foreshock and a long tail stretching out away from the Sun. However, starting in about 2010, data from Cassini and IBEX\footnote{\url{https://solarsystem.nasa.gov/missions/interstellar-boundary-explorer-ibex/in-depth/}} energetic neutral atom (ENA) measurements (5.2--55~keV), supported by SOHO and IBEX data, and from {\em in situ} Voyager observations, ``{\em strongly suggest a diamagnetic bubble-like heliosphere with few substantial tail-like features}'' \cite{Dialynas2017}. The heliosphere responds promptly, within $\sim$2--3 years, to outward-propagating solar wind changes in both the nose and tail directions.

{\bf\em Science objectives}: The long-accepted view of the shape of the heliosphere is that it is a comet-like object with a long tail opposite to the direction in which the solar system moves through the local ISM (LISM). The solar magnetic field at a large distance from the Sun is azimuthal, forming a spiral because of the rotation of the Sun. The traditional picture of the heliosphere as a comet-like structure comes from the assumption that, even though the solar wind becomes subsonic at the termination shock as it flows down the tail, it can stretch the solar magnetic field.  Based on magneto-hydrodynamic (MHD) simulations, Opher {\em et al.} \cite{Opher2020} established that the twisted magnetic field of the Sun confines the solar wind plasma and drives jets to the north and south very much like some astrophysical jets with the tension force being the primary driver of the outflow.

The shape of the heliosphere and the extent of its tail are thus subject to debate. According to the new model \cite{Opher2020}, as the Sun moves through the surrounding partially ionized medium, neutral hydrogen atoms penetrate the heliosphere, and through charge exchange with the supersonic solar wind, create a population of hot pick-up ions (PUIs). The termination shock crossing by Voyager 2 demonstrated that the heliosheath (the region of shocked solar wind) pressure is dominated by PUIs. However, the impact of the PUIs on the global structure of the heliosphere has not been explored. The new model \cite{Opher2020} reproduces both the properties of the PUIs, based on New Horizons observations, and the solar wind ions, based on the Voyager 2 spacecraft observations as well as the solar-like magnetic field data outside the heliosphere at Voyagers 1 and 2. It is therefore crucial to revisit LISM with new, modern {\em in situ} observations, which will be crucial to distinguish among the existing theories and to understand the physical picture of this region.

The main science questions that can be answered with sailcraft sent to various directions include:
\begin{inparaenum}[1)]
\item How does the solar wind interact with the ISM and how does this relate to the interaction of other stars with their interstellar surroundings and formation of stellar astrospheres?
\item How does this interaction lead to the observed complexities of the three-dimensional structure of the heliosphere?
\item What is the nature of the termination shock? What is the nature of the processes that govern formation of the heliosheath? What are the properties of the heliopause transition region?
\item How does the heliosphere affect the properties of the very local ISM, and how do they relate to the pristine ISM?
\end{inparaenum}

{\bf\em Mission design and requirements}: The shape of the heliosphere and the extent of its tail are subject to debate and the new model of the heliosphere---roughly spherical with a radius of $\sim$100~AU---needs confirmation. Of course, every mission out to $>$100~AU will test it, but a series of paired missions (nose and tail, and in perpendicular directions) would provide a substantial improvement in our understanding of ISM/solar wind interactions and dynamics. High-velocity, low-cost sailcraft could probe these questions related to the transition region from local to pristine ISM sooner and at lower cost than competing mission concepts. Since the exact trajectory is not that crucial, this would also provide excellent opportunities for ad hoc trans-Neptunian object flybys.

{\bf\em Instrumentation}: For more than a decade, the technique of ENA imaging has been brought to bear on mapping the structure of the outer heliosphere from vantage points relatively close to the Sun. As of 2022, a total of only nine dedicated ENA detectors had been flown. Space plasma observation using ENA imaging is an emerging technology that is finally coming into its own \cite{Gruntman1997}. Several improvements are still needed to perfect the technique, but the already available instruments may be flown on Sundivers.  Current measurements and modeling of the outer heliosphere put important constraints on its shape, though we are still not able to say unambiguously whether it is cometary, bubble shaped or `croissant-like'; and whether it is closed or open, or what processes are operating at different parts of its boundary. Making ENA measurements from beyond the heliopause will allow us to answer these questions directly.

{\bf\em Technology readiness}:
This mission will require transit velocities of 7--10~AU/yr. New sail materials capable of reaching close perihelion are needed. In addition, as the mission must reach regions farther than 100~AU, on-board power is the main on-board capability that must be developed. This mission objective is suitable for Phase II of Sundivers (see Sec.~\ref{sec:sdconcept}).

\subsection{The interstellar ribbon}
\label{sec:ribbon}

\begin{figure}
\includegraphics[width=\linewidth]{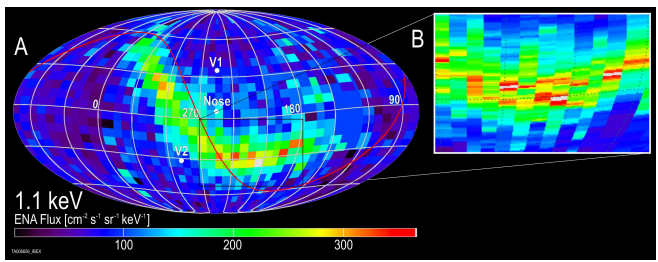}
\caption{\label{fig:IBEX}IBEX ENA Ribbon. A closer look suggests that the numbers of ENAs are enhanced at the interstellar boundary. A Sundiver spacecraft will go through this boundary as it travels to the ISM. Credit: SwRI.}
\end{figure}

Access to the outer solar system provides unprecedented opportunities to measure properties of the interplanetary dust (IPD) clouds \cite{Fuselier2009,Schwadron2019,Zirnstein2019}.  Complementary global maps from IBEX and Cassini/INCA, from the imaging of ENAs, show an unpredicted large-scale and narrow ENA structure first detected by IBEX, described as a ``ribbon'' (IBEX) or ``belt'' (Cassini) of ENA emissions from the outer heliosphere, apparently ordered by the local interstellar magnetic field (Fig.~\ref{fig:IBEX}).
This feature is narrow ($\sim$20${}^\circ$ average width) but long (extending over 300$^\circ$ in the sky). It is observed at ENA energies of 0.2--6~keV. The ribbon parallax reveals a distance of $140^{+84}_{-38}$~AU \cite{Swaczyna-etal:2016}. It is generally thought to be a feature related to the interaction of the heliosphere and the local ISM.

{\bf\em Science objectives}: How and where the ribbon and belt are produced is a subject of much debate. Other questions include:
\begin{inparaenum}[1)]
\item What is the source of the strong temporal variation?
\item Why do ENA maps of the heliosphere's tail region also show unexpected depletion areas?
\item What causes the dynamic role of the interstellar magnetic field in shaping the outer heliosphere to be stronger than expected?
\item Why does the magnetic field measured by Voyager 1 in the LISM not have the direction inferred from various remote sensing observations, including the bright ribbon?
\end{inparaenum}

The most recent measurements from Voyager 1 show that the influence of solar wind extends farther into the local ISM than expected, but it is not known why or how far. The unexpected results from Voyager, IBEX, Cassini, and other observations demonstrate the limitations in our understanding of the interactions between stars and the interstellar environment and the need to revisit that region with modern instrumentation, sensitive to the low magnetic fields in the heliosheath, as well as to measure pick-up ions and to measure the anomalous, galactic cosmic rays (ACR/GCR). These measurements are crucial to sorting out the different scenarios and improve our understanding of our place in the Galaxy.

{\bf\em Mission design and requirements}: To address the above science objectives, {\em in situ} measurements are needed of all components:
\begin{inparaenum}[1)]
\item ISM and solar wind plasma electrons, ions, and neutrals;
\item solar and ISM magnetic fields, electromagnetic waves and turbulence;
\item Energetic particles and cosmic rays; and
\item Dust.
\end{inparaenum}
Depending on the component, the energy distribution functions or elemental and isotopic compositions need to be measured. To obtain remote measurements relating to the structure and dynamics of the heliosphere and ISM, observations such as ENA imaging and Lyman-alpha observations are needed.

{\bf\em Instrumentation}: A mission would have to provide {\em in situ} measurements of particles and fields, especially magnetic fields, energetic particle detectors (including neutral particle detectors, suprathermal particles and cosmic rays), plasma wave and thermal plasma instruments and dust detectors. There would be great value in a series of probes to study the ribbon at different locations. It is such a large-scale feature that its exploration could be combined with flybys of a trans-Neptunian object or other solar system objects. However, there are no systematic attempts to find such dual-purpose trajectories.

{\bf\em Technology readiness}: New thermally resistant reflective materials, thermally protective sailcraft design and avionics and autonomy are needed for the close perihelion required for high transit velocities. On-board power is the main capability that must be developed. Missions addressing these objectives are suitable for Phase II of Sundivers.

\subsection{Studies of the pristine interstellar medium}
\label{sec:pism}

The ISM immediately outside our solar system is the closest example of cosmic {\em terra incognita}. It deserves a special place in our exploration agenda. Places we have explored directly with probes so far have been limited to solar system objects and the interplanetary medium which is dominated by phenomena originating from our Sun \cite{ISM2015}. The gas and dust drifting among the stars, the ISM, is found throughout our Galaxy and is an integral part of energy flows and the material cycles defining the ecology of all galaxies. The ISM is the repository of the raw materials from which new stars form. It is replenished by stars at the end of their evolutionary cycles.

The LISM has been sampled by the two Voyager spacecraft, but only in the heliosphere-ISM transition region. Our knowledge of the local ISM beyond that region is from optical absorption line measurements from nearby stars. These measurements reveal that the solar system appears to be on the boundary of two ISM clouds: the G cloud, which lies towards the galactic center and includes the 3 stars of the Alpha Centauri system; and the Local Interstellar Cloud (LIC), which lies mostly in the opposite direction. At present, it is not clear if the solar system is in the G cloud, the LIC, or in a transition region between the two clouds, nor are there any good models of such a transition region. This is a difficult and important problem that can best be resolved by {\em in situ} exploration using multiple probes, including probes both toward and away from the galactic center.

{\bf\em Science objectives}: The LISM determines the boundary conditions that dictate the interaction of the Galaxy with the Sun. This interaction defines the shape and extent of the heliosphere. Given the relatively large variability in ISM density and the relatively small variability in solar emissions, the ISM dominates the general structure of the heliosphere, which acts as a boundary, shielding us from the ISM and limiting the flux of cosmic rays to the solar system. It is therefore of great importance to fully characterize its properties and to conduct in-situ measurements of grains/dust. The ISM should be sampled beyond the LISM, where it is unperturbed by any interaction with the Sun. This pristine ISM is located beyond the bow wave/shock, at $>500$~AU from the Sun. The heliosphere acts as a filter, and there are components of the ISM that make it into the inner solar system. However, most of the material that makes up the ISM can only be sampled in its pristine form beyond the heliosphere.

In this vast volume of space, immediately beyond the heliosphere, where we find the pristine ISM, we also find the closest stars and planetary systems. Thus, even in our most immediate cosmic neighborhood, perhaps within a sphere of 10 pc, we find a complex morphology of ISM clouds, hundreds of stars, dozens of known exoplanets, and a handful of astrospheres (structures analogous to the heliosphere around other stars). Accounting for the fact that stars with winds, orbited by planets, and adrift in the ISM are ubiquitous, the study of our own heliosphere and its interaction with the Galaxy will help us understand other analogous structures. Direct investigations of the LISM will provide critical information for understanding the chemical evolution and mixing of matter within galaxies (e.g., neutrals, ions, isotopes, molecules, dust).

{\bf\em Mission design and requirements}: Sampling each matter constituent will allow measuring fundamental elemental abundances, as current line-of-sight observations suffer from limited observational capability for some elements. Therefore, {\em in situ} measurements are vital as they provide access to data that will have a critical impact on evaluating degeneracies or uncertainties in the long-sightline average. As with studies of the heliosphere, as a precise trajectory is not essential, so missions can be combined with a flyby of a suitably located asteroid or trans-Neptunian object.

{\bf\em Instrumentation}: Sundivers may provide unique data on the properties of the pristine ISM by conducting observations of ENAs and interstellar neutrals (ISN). Observations of neutral atoms can be grouped into several types of ENA and ISN sources. The instruments should be able to observe neutral atoms over the energy range from 10~eV to roughly 6~keV and also both ISN atoms (species-dependent energies from 10~eV to 0.8~keV for ram observations in Earth orbit), and ENAs from the heliosheath (the plasma region between the solar wind termination shock and the heliopause) and from the perturbed ISM outside the heliopause \cite{Galli2022}. The already available instrumentation is low-weight and low-power, simplifying accommodation on a Sundiver vehicle.

{\bf\em Technology readiness}: New thermally resistant reflective materials, thermally protective sailcraft design and avionics and autonomy are needed for the close perihelion required for high transit velocities. On-board power is the main capability that must be developed. Missions addressing these objectives are suitable for Phase II-III of Sundiver program.

\section{Planetary science by fast transit}
\label{sec:planetary}

One of the most exciting applications of the new mission architecture is fast and frequent exploration of the outer solar system. Since the beginning of the Space Age, only six spacecraft have ventured beyond Jupiter. Four of these launched before 1980. We have sent fifty missions to Mars but we have flown by the two most distant planets in our solar system, Uranus and Neptune, only once each. These brief flybys took place almost forty years ago. There have been proposals to return but, as recently as 2019, NASA selected two Venus missions over Neptune- and Io-bound missions as part of its 2019 Discovery Program. While Venus is an understudied world, in need of renewed attention, it is the cost and time commitments that seem to make outer solar system destinations unpopular\footnote{\url{https://www.nasa.gov/press-release/nasa-selects-four-possible-missions-to-study-the-secrets-of-the-solar-system}, \url{https://www.nasa.gov/press-release/nasa-selects-2-missions-to-study-lost-habitable-world-of-venus}}. However, this dearth of outer solar system missions means that even low-cost missions with simple instrumentation will provide significant science payoffs; the icy moons of the gaseous planets are ideal places to look for complex organic molecules, or even life.

\subsection{Study of hard-to-reach asteroids}
\label{sec:astroids}

An excellent use case for early missions by high-thrust sailcraft is to reach bodies that are either too hard to reach with chemical propulsion (as the required $\Delta v$ may be too large), or where there is a limited time to launch the mission. Asteroids are very heterogeneous in composition, size, and origin. Unlike larger bodies, the surfaces of most asteroids have not been altered by weather or geological processes. They are considered the fundamental building blocks of planets and moons, providing a pristine view into the earliest period of the solar system before the formation of the major planets and moons. It is also possible that asteroid impacts may have brought water and organics to the early Earth. In addition, nation states and private companies are becoming interested in asteroids as sources of valuable resources and rare minerals. While asteroid mining might still be a few years away, we first need to characterize the more than 30,000 asteroids, and asteroid prospecting is likely to become an important new class of space missions in the coming years. Unlike with traditional propulsion, sailcraft are able to visit and characterize multiple asteroids with simple instruments. Furthermore, planetary protection requires close monitoring of potentially hazardous asteroids. If an asteroid is determined to be a threat, immediate action may be required to deflect a potentially catastrophic trajectory. Not all potentially harmful asteroids are easy to reach using traditional propulsion. Solar sailing provides an inexpensive, viable alternative.

{\bf\em Science objectives}: Since asteroids are small and have low albedo, it is important to observe them {\em in situ}. Important information about an asteroid's size, mass, shape, and composition can be acquired with relatively simple instrumentation. This data can be used to characterize and categorize asteroids, precisely measure their trajectories, and improve our understanding of their environments. Data to be acquired includes orbit and exact position in space; precise shape; rotational properties; spectral class; local dust and debris fields; morphology; composition, and magnetosphere\footnote{\url{https://www.jpl.nasa.gov/missions/near-earth-asteroid-scout-neascout}} among other topics.

{\bf\em Mission design and requirements}: There are thousands of asteroids worth exploring and each will require a slightly modified mission design. Solar sails offer a means to explore all of them; sundiver missions ideally suited to reach more distant asteroids and some on more atypical orbits, and standard lightsail missions that do not need to sundive for many of the close asteroids. Indeed, two of the latest major lightsail missions (NASA's NEA Scout and JAXA's OKEANOS \cite{Okada:2019}) both have asteroid targets. Some top candidates for early investigation by Sundivers are:
\begin{inparaenum}[1)]
\item 2010 TK7 \cite{Connors2011};
\item C/2014 UN271 \cite{Hui2022};
\item (594913) \'Ayl\'o\'chaxnim; and
\item (1566) Icarus
\end{inparaenum}
\cite{Campbell1983,Shapiro1971}, each is described briefly below. We emphasize that the exploration of any of these targets can be easily accommodated during the early phases of the Sundiver concept development.

\begin{itemize}
\item 2010 TK7 is the largest Earth trojan. Earth trojans are easy to reach energetically, but harder in a chemical mission (they are typically $\sim$1~AU away). These objects are thus going to be good sailcraft targets and would serve as good initial ``easy deep space'' targets.
\item Icarus is a (1.4~km${}\times{}$1.2~km) potentially hazardous asteroid (PHA) on a highly elliptical orbit ($e = 0.827$), which crosses the orbits of Mercury, Venus, Earth, and Mars, and would be very energetically expensive to reach using chemical propulsion. The mission profile would include a rendezvous with Icarus at low velocity and the delivery of a tracking beacon or transponder onto the asteroid's surface, together with photography and spectroscopy at close range.
\item 594913 \textsc{\char13}Ayl\'o\textsc{\char13}chaxnim is an asteroid with an orbit entirely interior to Venus, has an aphelion distance of 0.65~AU, is 2~km in diameter and red in color \cite{Bolin2022}. The detection of such a large asteroid inside the orbit of Venus is surprising, given their rarity according to near-Earth asteroid population models. A mission could rendezvous with this asteroid, imaging it and performing close-range spectroscopy.
\item C/2014 UN271 is by far the largest known Oort cloud object, with an effective diameter of $137\pm 15$~km \citep{Bernardinelli-Bernstein:2021}. UN271 will never come closer than the orbit of Saturn, with perihelion passage in 2031 and ecliptic passage in 2033. A New Frontiers or Flagship class mission launched in the late 2020s could easily reach UN271 at the time of ecliptic passage, but no such mission is in preparation, leaving sailcraft as the only feasible means of reaching it.
\end{itemize}

{\bf\em Instrumentation}: Techniques that are important to characterize these targets include
\begin{inparaenum}[1)]
\item Imaging that will tell us about the object's geological history; its shape can give us information on density and structure;
\item Orbital analysis will yield density and density distribution by mapping the object's gravitational field;
\item Magnetometry will probe for the presence of water if there is a varying background field; and
\item Mass spectroscopy and chemical analysis tools will provide information on the atmospheric composition, surface chemistry, and potentially  the origin of constituent materials.
\end{inparaenum}

These types of investigations may be done with a suite of instruments that include a high-resolution visible light imager, an optical and near-IR imaging spectrometer and a thermal IR spectrometer, like those flown on the Lucy mission\footnote{\url{https://www.nasa.gov/mission_pages/lucy/overview/index/}}. The utility of such investigations was proven for the upcoming Psyche mission\footnote{\url{https://psyche.asu.edu/}} which includes:
\begin{inparaenum}[1)]
\item a multispectral imager needed to obtain high-resolution imagery, with filters that allow this instrument to identify the metals and silicates or rocky materials that make up the asteroid Psyche, and powerful twin cameras to gather data on the asteroid's geology, composition, and topography, with one of the cameras assisting with optical navigation;
\item a gamma ray and neutron spectrometer to unmask the chemical elements on asteroid Psyche's surface: iron, nickel, silicon, and oxygen; and
\item a magnetometer to reveal the asteroid's history and composition by measuring its magnetic field.
\end{inparaenum}
Miniaturized instruments that can provide similar science measurements will be ideal science payloads on a Sundiver mission to these bodies.

{\bf\em Technology readiness}: Reaching asteroids is a natural application for solar sailing. The overall technology readiness for such missions is quite high, requiring no major technology development \cite{Garber2022b}. Such missions may be implemented in Phases I--II of the overall Sundiver program.

\subsection{Probing the plumes on outer solar system moons}
\label{sec:plumes}

Our knowledge of the ongoing processes in the planetary systems residing in the outer solar system was advanced considerably by the 13-year long, in-depth, comprehensive exploration conducted by the Cassini--Huygens mission at Saturn.  In particular, Cassini's study of the small moon, Enceladus, by both remote sensing and {\em in situ} investigations demonstrated its status as a prime candidate for astrobiological study and the search for evidence of life.  It is home to a subsurface global water ocean (e.g., \cite{Thomas-etal:2016}) that is likely long-lived (e.g., \cite{Fuller-etal:2016,Lainey-etal:2017}) and vents through 4 long, prominent fissures in the moon's south polar terrain ice shell in the form $\sim100$ discrete geysers of vapor and icy particles, with faint sheets of material in between \cite{Porco-etal:2014,Teolis-etal:2017}.

The large plume formed by both discrete and fissure eruptions---both vapor and solids---extends hundreds of kilometers above the moon's south polar terrain and is variable in strength but persistent (e.g., \cite{Nimmo-etal:2014}). Cassini's fly-throughs of this plume allowed the spacecraft's {\em in situ} instruments to determine its contents: salty water, with trace amounts of simple and large complex organic molecules and other biologically significant compounds (e.g., \cite{Postberg-etal:2009,Waite-etal:2009,Kopparla-etal:2016,Waite-etal:2017,Postberg-etal:2018}).  Evidence has also been found in the plume solids of compounds that are best-explained by seafloor hydrothermal activity with alkaline pH values and temperatures \cite{Hsu-etal:2015}, similar to that of terrestrial off-axis low temperature hydrothermal zones in the mid-Atlantic.
The recent report on the detection of phosphates in the plume originating from Enceladus?s ocean raises the possibility that life could exist in the moon?s plume-forming ocean waters \cite{Postberg-etal:2022,Postberg-etal:2023}.

Because its plume is readily accessible and always present, Enceladus is an excellent target for Sundiver missions. (Although there are reports of plumes erupting from the surface of Europa, that conclusion is still debated. For this reason, we concentrate our discussion here on Enceladus, recognizing that the same principles could be applied to Europan plumes should they be ultimately confirmed.)

{\bf\em Science objectives}: Sailcraft flying through the Enceladus plume could carry light-weight instrumentation designed to investigate specifically whether or not the moon's ocean contains signs of biological activity, something Cassini was not equipped to do.  In addition to the usual approaches of  identifying and measuring the characteristics of particular organic compounds, or searching for enantiomeric excess which, if large enough, would be a strong indicator of life, it would be possible to employ more modern, yet unutilized approaches.  One  is Assembly Theory \cite{Sharma-etal:2022}, which proposes that the presence of life in any environmental sample---in this case, a plume sample---could be ascertained by determination of a simple metric based on complexity theory: If the sample measurement exceeds the threshold, life is present. Another new approach is a search for a polyelectrolyte: a particular molecular structure that, it is proposed, any genetic molecule must possess \cite{Schrodinger:1944,Benner:2017}.  Both of these approaches are completely chemically agnostic and do not rely on the living system having terrestrial-like biochemistry, making these methods powerful hedges against our assumptions of Earth-based biochemistry.

{\bf\em Mission design and requirements}: Clearly, {\em in situ} investigations of the plumes are critical. We envision the use of 2--3 solar sailing microsats accelerated to 5--7~AU/yr. Sailcraft could fly through the plume as Cassini did.  However, the data must be taken at relative speeds $<5$~km/s to avoid vaporizing the very molecules we are trying to collect. One sailcraft could carry one suite of instruments. The other two would have propulsion modules to be used for a negative Oberth maneuver at Saturn and Rhea (note that doing so at Titan may result in organic molecules from Titan's atmosphere, contaminating the equipment before getting to Enceladus), needed to slow the microsat to $<1$~AU/yr as required for reliable sampling of molecules from a plume during a single flyby.

Another mission type may rely on in-flight aggregation \cite{Helvajian2022}, which may be needed to allow for orbital capture. For that, after perihelion passage and while moving at 5~AU/yr ($\sim$25~km/s), the microsats would perform in-flight aggregation to make a fully capable smallsat to satisfy conditions for {\em in situ} investigations. One such important capability may be enhanced on-board propulsion capable of providing the $\Delta v$ needed to slow down the smallsat. In this case, before approaching Enceladus, the spacecraft reduces its velocity by 7.5~km/s using a combination of on-board propulsion and gravity assists. Moving in the same direction with Enceladus (which orbits Saturn at 12.6~km/s) it achieves the conditions for {\em in situ} biomaterial collection.

To target the plumes, we could deploy several sailcraft (sterilized in advance to avoid planetary contamination) to sample multiple locations to detect the presence of organic molecules in the plumes. Swarms of inexpensive small sailcraft could provide opportunities to significantly enhance infrequent interplanetary missions with, e.g., landers or sacrificial satellites, and networks of small satellites that could enable missions to these unique objects to detect and study life that may exist on these bodies in the outer solar system. Swarms could perform initial observations of these watery worlds and inform the development of future systems dedicated for their exploration.

{\bf\em Instrumentation}: At a minimum, a Sundiver mission to Enceladus would need a mass spectrometer and a camera for, respectively, {\em in situ} investigation of the plume contents and imaging for both navigation and context.  Previous well-developed, non-Sundiver mission concepts for Enceladus life-detection (e.g., \cite{MacKenzie-etal:2021}) have considered specific science instruments, such as both gas and ice-particle mass spectrometers; a microfluidics electrophoresis ``lab-on-a-chip'' organic chemical analyzer that can detect in very small quantities amines, amino acids, and even amino acid chirality biomarkers; microscopes to detect cells and/or organisms as well as determine the morphology of ice particles; and an instrument to search for molecules of polyelectrolyte structure, like that of DNA.  If a mission to Enceladus includes multiple Sundivers, then the required instruments can, in principle, be split among them to achieve the science objectives.

{\bf\em Technology readiness}:  For {\em in situ} investigations, the sailcraft would have to slow down to below 4~km/s. This can be done by in-flight (as opposed to Earth-orbiting or cislunar) autonomous assembly of a large spacecraft that is built from modules, separately delivered in the form of microsats ($<$20~kg), where each microsat is placed on a fast solar system transit trajectory via solar sail propulsion to velocities of $\sim$10~AU/yr. The benefits of in-flight aggregation relate to the propulsion module that would allow slowing down the probe before it enters the plumes of Enceladus. Thus, solar sailing shortens the flight time to destination, while in-flight aggregation reduces the relative speed to take valuable data.  The same critical technology elements that are required for the missions to distant regions (power sources\footnote{A promising technology is being developed by the Aerospace Corporation, see Atomic Planar Power for Lightweight Exploration (APPLE): \url{https://www.nasa.gov/directorates/spacetech/niac/2022/Atomic_Planar_Power/}}, advanced materials, etc.) are important. Missions to probe the plumes of Enceladus are suitable for Phase II of Sundivers and may be implemented at the beginning of the next decade.

\subsection{Molecular biosignatures in the solar system}
\label{sec:mol-bio}

Typically, the search for life in our solar system has involved picking a single target, and then sending a highly capable mission to test for specific biomarkers or the ingredients necessary for life. To date, Mars has been the primary target of such missions although remote observations of other bodies have picked up possible traces of non-equilibrium molecules or organics that could indicate life. With the ability to send multiple fast, cheap probes all over the solar system, a campaign to fully ``biomap'' the solar system becomes a possibility. Rather than investing all our resources in exploring one planetary target for life detection, the search for life will be a lot more effective if we can send a dozen probes to different planetary bodies to provide a complete map of the molecular biosignatures in our solar system.

{\bf\em Science objectives}: Given the many unknowns regarding the surface environment and chemical composition of our planets in the solar system, the development of a complexity-first observation approach could be experimentally very powerful. This is because the observations of potential biomarkers like methane, oxygen, or phosphine, would need to be traced to a source. The problem with these molecules is that they are very simple and carry no information regarding their origin.

The Sundiver probes could be equipped with a suite for infrared, Raman, and fluorescence spectroscopy, for remote detection of small molecules in the atmosphere, and possibly a mass spectrometer and an optical instrument for deployment beyond Earth, to study other targets including Venus, Mars, Europa, Titan, Enceladus. The primary objective would be to detect complex molecules and use the results in a probabilistic framework to be applied to the life detection.

The spectroscopic payload could be used not only to probe the identities and concentrations of simple bioindicators of the biomarkers in the atmosphere, but also look for more complex spectroscopic signatures. The use of fluorescence IR emission might serve as a new remote technique to allow the separation of various species in the atmosphere to probe the intrinsic complexity of the molecules present rather than recording spectra from complex mixtures. A fragmentation mass spectrometer would be a powerful tool if deployed in the atmosphere for direct sampling since the complexity of molecules in the atmosphere could be probed directly (again, we would need to consider slowing down the probe before taking the sample). Furthermore, an optical microscope with a microfluidic sample insertion system could be used to look at the morphology of particles found in the atmosphere between 500~nm and 100~$\mu$m in size. A miniaturized spectroscopic workstation could be used to probe the chemistry of the particles also.

{\bf\em Mission design and requirements}:  Investigations of biomarkers may be done remotely by a spectroscopy suite onboard a Sundiver. An IR-fluorescence system, suitable for mapping molecular complexity \cite{Marshall-etal:2021,Liu-etal:2021} remotely, may be included for spectroscopic investigations.  These missions must be flown in proximity of the target body to allow for good instrument pointing.

{\bf\em Instrumentation}: A mission framework should consider including the following instruments:
\begin{inparaenum}[1)]
\item Raman and IR spectrometer;
\item Fluorescence spectrometer;
\item Fragmentation mass spectrometer;
\item Optical microscope for image analysis.
\end{inparaenum}
Depending on the target and the choice of detection, a single orbital probe or possibly a duplicate of the detection system on a lander. The probe size could be very small with minimal power requirements if an IR detection system is used. If we are to use a mass spectrometer, significantly more mass and power would be required. There are a range of flight-ready mass spectrometers available (cf. units on Mars and in manufacture for DragonFly mission\footnote{\url{https://dragonfly.jhuapl.edu/}} to Titan \cite{Grubisic-2021}.) The development of a tandem machine would reduce the science risk but increase the technical risk. In terms of spectroscopic systems, we are exploring how to adapt current spectroscopic packages deployed in low earth orbit and validate them. An orbitrap mass spectrometer would be required, or that would lower risk considerably or might be required. There exists a space orbitrap consortium, potentially providing an opportunity to collaborate and reduce further development risk significantly.

{\bf\em Technology readiness}: Although the transit speed is important, the main benefit of Sundivers is their precision navigation that will allow delivery of the {\em in situ} instruments to the target regions. The same critical technology elements that are important for the missions to the distant regions of the solar system are important here. Accordingly, these missions are suitable for Phase II of Sundiver program.

\subsection{Fast flybys of Uranus and Neptune}
\label{sec:UrNep}

Since Voyager 2 performed Uranus and Neptune flybys between 1986 and 1989, no other mission has taken place to study the most distant planets in our solar system. As a result, Uranus and Neptune are planetary systems about which we know the least. Even the simplest mission to either planet would offer an outsized impact on our understanding of the planets that make up quarter of our solar system \cite{Guillot:2022}.

{\bf\em Science objectives}: Since so little is known about the ice giants, numerous questions remain to be answered. Ice giants are the only unexplored class of planets in the solar system, and their exploration is key to understanding how the solar system formed and evolved. A mission would provide an opportunity to study their rings, atmosphere, auroras, and moons. Neptune's moon Triton is a major moon in the solar system in retrograde orbit, and is believed to be a captured dwarf planet from the Kuiper Belt. Triton is also of interest because little is known about the origin and prevalence of the plumes that were imaged by Voyager 2. In fact, Triton has been identified as the highest priority candidate ocean worlds by the NASA Outer Planets Assessment Group (OPAG) Roadmap to Ocean Worlds \cite{Hendrix-etal:2019}. Four prime areas of interest are the ice giant's
\begin{inparaenum}[1)]
\item atmospheres;
\item interiors, magnetospheres and aurorae;
\item ring systems; and
\item satellite systems.
\end{inparaenum}

Concerning ice giant atmospheres, the following questions arise:
\begin{inparaenum}[1)]
\item What are the dynamical, meteorological, and chemical impacts of the extremes of planetary luminosity?
\item What is the large-scale circulation of ice giant atmospheres, and how deep does it go?
\item How does atmospheric chemistry and haze formation respond to extreme variations in sunlight and vertical mixing, and to the influence of external material?
\item What are the energy sources responsible for heating their middle and upper atmospheres?
\item How do planetary ionospheres enable the energy transfer that couples the atmosphere and magnetosphere?
\end{inparaenum}

When considering ice giant interiors, magnetospheres and aurorae, we are interested in learning the following:
\begin{inparaenum}[1)]
\item How did the ice giants first form, and what constraints can be placed on the mechanisms for planetary accretion?
\item What is the role of giant impacts in explaining the differences between Uranus and Neptune?
\item What is the bulk composition and internal structure of Uranus and Neptune?
\item How can ice giant observations be used to explore the states of matter (e.g., water) and mixtures (e.g., rocks, water, H-He) under the extreme conditions of planetary interiors?
\item What physical and chemical processes during the planetary formation and evolution shape the magnetic field, thermal profile, and other observable quantities?
\item Is there an equilibrium state of the ice giant magnetospheres?
\item How do the ice giant magnetospheres evolve dynamically? viii) How can we probe ice giant magnetospheres through their aurorae?
\end{inparaenum}

The study of the ring systems will address the following questions:
\begin{inparaenum}[1)]
\item What is the origin and composition of planetary ring systems, and why are they so different?
\item How do the ring-moon systems evolve?
\end{inparaenum}

When satellite systems are addressed, we can learn answers to questions such as:
\begin{inparaenum}[1)]
\item What can the geological diversity of the large icy satellites of Uranus reveal about the formation and continued evolution of primordial satellite systems?
\item What was the influence of tidal interaction and internal melting on shaping the Uranian worlds, and could internal water oceans still exist?
\item What is the chemical composition of the surfaces of the Uranian moons?
\item Does Triton currently harbor a subsurface ocean and is there evidence for recent, or ongoing, active exchange with its surface?
\item Are seasonal changes in Triton's tenuous atmosphere linked to specific sources and sinks on the surface, including its remarkable plume activity?
\item Are the smaller satellites of Neptune primordial?
\item How does an ice giant satellite system interact with the planet's magnetosphere?
\end{inparaenum}

{\bf\em Mission design and requirements}: A mission to accomplish these objectives would involve an interplanetary probe performing a flyby of the target system. The following flyby opportunities are of interest when we investigate Triton, Neptune's only big moon: to find out if it is a captured Kuiper Belt Object (KBO); to confirm the existence of a suspected ocean on Triton \cite{Cochrane2022}; and to study its young surface, active plumes, thin atmosphere, and intense ionosphere.

{\bf\em Instrumentation}: To investigate Uranus and Neptune from a fast flyby, we would refly a New Horizons mission, adding a magnetometer to the science payload. In general, we need to select lightweight instruments. Instrument miniaturization is a key for Sundivers. Considering the set of instruments, a narrow-angle camera is top priority, followed by a magnetometer. Radio science investigations using X-band or Ka-band communication links are an important part of the science package.

A mission concept for outer planets by integrating an ultra-lightweight quantum dot-based spectral imager with the sailcraft is currently being developed \cite{Sultana-etal:2022}. These payloads would be printed directly on the solar sail, making them especially suitable within the tight mass constraints of solar sail missions.

{\bf\em Technology readiness}:  Critical technology elements important for the missions to distant regions (power sources, advanced materials, etc.) need to be developed. Such missions are suitable for Phase II of Sundivers.

\subsection{Kuiper Belt and Oort Cloud objects}
\label{sec:kbooco}

The Kuiper belt is a disc-shaped region beyond the orbit of Neptune, extending to 50 AU from the Sun. Since its discovery, the number of known Kuiper Belt Objects (KBOs) has steadily increased. More than 105 KBOs over 100 km in diameter are thought to exist. The recently discovered dwarf planets Haumea, Makemake, Eris, and Quaoar all provide interesting targets for exploration. These objects orbit the Sun at the very edges of the solar system at distances ranging 40--90 AU. Another interesting target would be Sedna, or other objects from the new family of extended scattered disc objects recently discovered with semimajor axes extending well beyond 100 AU.

{\bf\em Science objectives}: A mission reaching out to the outer solar system presents a unique opportunity to fly by a large KBO. Various KBO candidates were studied for a near-term mission with Makemake, Haumea, and Quaoar determined to be of high science value. Out of these three, Quaoar is the most studied and is one of the most interesting KBOs. It is undergoing a transition between the large, volatile-dominated, atmosphere-bearing planetesimal and a typical mid-sized, volatile-poor object. For most of its history, it had a methane atmosphere, but it is now in the last stages of losing it \cite{Arimatsu2019}. Most likely, its surface is patchy with methane frost, with the methane being mostly cold trapped near the poles or in craters. The processes related to atmospheric loss in the outer solar system are poorly known, so Quaoar offers an interesting opportunity to see the process in its late stages. Given its size, Quaoar may have ancient cryovolcanic flows on the surface offering clues on its history. To investigate these processes, we can make full global imaging in broadband colors, which may be achieved with a swarm of sailcraft. In addition, we may study Quaoar's surface using one of the sailcraft as an impactor. Imaging the crater would be an interesting probe into surface conditions. Plume spectroscopy could explore subsurface composition.

Key science questions include
\begin{inparaenum}[1)]
\item What is the fraction of cryovolcanic coverage?
\item When was the last activity?
\item What is the depth and coverage of methane?
\item What is the crater count and ages of frosty surfaces?
\item What is the spatial distribution of volatiles?
\item What is the mass ratio of Quaoar/Weywot?
\item Are there additional moons?
\item What is Quaoar's interior structure?
\item What is the structure of Quaoar's atmosphere?
\end{inparaenum}
Similar questions could also help study Haumea and Makemake.

{\bf\em Mission design and requirements}: The missions discussed above are destined to distant regions of the other solar system. They will have to be able to perform precision flybys of the target body. These missions would therefore require more sophisticated guidance, power, and communications capabilities. Imaging equipment on those sailcraft needs to be able to compensate for the high relative velocity with respect to the target body. In addition, swarms of Sundiver sailcraft may be used to carry out these investigations. As such, these missions are good candidates for later iterations of the Sundiver spacecraft, those that will be available in Phase III of the Sundiver program.

{\bf\em Instrumentation}: These missions will require instrumentation similar to those suggested for the investigations of Uranus and Neptune and discussed above. In general, we would want to have a narrow-angle camera and a magnetometer. As before, radio science investigations may be done using communication links in the X or Ka-band.

{\bf\em Technology readiness}: New thermally resistant reflective materials, thermally protective sailcraft design and avionics and autonomy are needed for the close perihelion required for high transit velocities. On-board power is the main capability that must be developed. Missions addressing these objectives are suitable for Phase II of Sundivers.

\subsection{Probing Planet 9}
\label{sec:planet9}

The clustering of orbits for a group of extreme trans-Neptunian objects suggests the existence of an unseen planet, the so-called {\em Planet~9}, with a mass
$\sim6^{+2.2}_{-1.3} M_\oplus$, semi-major axis of ${380}_{-80}^{+140}$ AU, inclination of $16 \pm 5^\circ$, and  perihelion of ${300}_{-60}^{+85}$ AU
\cite{Batygin2019,brown2021orbit}. Direct imaging searches have not yet detected this {\em Planet~9}. Recently \cite{Witten2020} it was suggested to use of a relativistic sailcraft, envisioned by the Breakthrough Starshot program, to indirectly probe {\em Planet 9} through its gravitational influence on the probe's trajectory. A similar proposal was made earlier to measure the mass of planets via interferometry by an array of Starshot-type spacecraft \cite{Christian2017} and to probe {\em Planet 9} \cite{Loeb2019,Heller-etal:2020}. Also, the transverse effect of gravity was considered in \cite{Lawrence2020} where the angular deflection of the probe's trajectory was derived to be $\sim$$10^{-9}$~rad. It was argued that an angular deflection of this magnitude can be measured with an Earth-based or near-Earth based telescope suggesting that their method is better than measuring the time delay because the transverse effect is permanent, whereas the time delay is only detectable when the spacecraft passes close to {\em Planet 9}.

{\bf\em Science objectives}: Finding a new major planet in the solar system would be an important discovery.  We would seek to determine its mass, size, structure and composition. As the mass of Planet 9 is estimated to be around $\sim6^{+2.2}_{-1.3} M_\oplus$ \cite{brown2021orbit}, it would provide insights into a class of planets with a size between Earth and Neptune \cite{hibberd2022can}. By imaging {\em Planet 9}, we could obtain spatially resolved data of its surface and interior, learning on its history. In particular, the question of how Planet 9 formed and whether it was assembled {\em in situ} in the solar system or was captured could be answered. It might even be possible to search for biosignatures.

{\bf\em Mission design and requirements}: As {\em Planet 9} is expected to orbit the Sun at a very large distance, we cannot rely on chemical propulsion to explore the object as it would take $>40$ years to reach it \cite{hibberd2022can}. Solar sailing smallsats on fast trajectories could conduct initial exploration with trip times of $~20$ years.  Once the planet is found, the next step would be to deploy Sundivers toward it. This could be a single scout mission, or it could be a swarm of scouts. If the orbit of {\em Planet 9} is not known with certainty, Sundivers could be used to refine the orbital model. For this, we would deploy microsats with precision navigation capability. Using their navigational data, we can model the source of gravitational perturbations and establish the precise orbit of {\em Planet 9} in the outer solar system.

{\bf\em Instrumentation}: To study {\em Planet 9}, we would need an imaging camera capable of operation in the low-light environment from a fast-moving platform. We could also carry an impactor and a spectrometer that could be used to study the surface composition of this celestial body. In addition, the spacecraft must be able to support high precision navigation capability for trajectory control and gravity research.

{\bf\em Technology readiness}: New thermally resistant reflective materials, thermally protective sailcraft design and avionics and autonomy are needed for the close perihelion required for high transit velocities. Precision navigation capability is needed. On-board power is the main capability that must be developed. These missions may be implemented toward the end of Phase II or early in Phase III of the Sundiver program.

\section{Astrophysics investigations}
\label{sec:astro}

Astrophysics is also poised to greatly benefit from the Sundiver paradigm. Fast access to distant regions of the solar system allows for a number of exciting science investigations that may answer a set of long-standing questions as well as to probe recent discoveries.

\subsection{Observing Earth as an exoplanet}
\label{sec:earth-exo}

Over the past two decades, thousands of extrasolar planets have been discovered, almost all of them extremely different from any world in our own solar system. The recent NAS Decadal Survey in Astrophysics \cite{NASEM2021} had emphasized the need for the interlinked studies of stars, planetary systems, and our solar system. The survey has identified the priority science area of Pathways to Habitable Worlds with the goal of trying to discover worlds that could resemble Earth and answer the fundamental question: ``Are we alone?''

It is remarkable that our generation could realistically discover evidence of life beyond Earth \cite{Seager2014}. The magnitude of the question of whether we are alone in the Universe, and the public interest therein, opens the possibility that results may be taken to imply more than the observations support, or than the observers intend. As life-detection objectives become increasingly prominent in space sciences, it is essential to consider how uncertainty in separate lines of evidence propagates into overall confidence.

Several exoplanetary missions (Kepler\footnote{\url{https://www.nasa.gov/mission_pages/kepler/overview/index.html}}, TESS\footnote{The Transiting Exoplanet Survey Satellite (TESS): \url{https://www.nasa.gov/tess-transiting-exoplanet-survey-satellite}}, JWST\footnote{The James Webb Space Telescope (JWST): \url{https://www.nasa.gov/mission_pages/webb/main/index.html}}) provided us with detailed spectra to understand the conditions and compositions of exoplanetary atmospheres, which may ultimately lead to detectable indicators of life. In order to interpret these observations, we need detailed models of what the atmosphere of an inhabited planet might look like. The best available source of ``ground truth'' for such models would be satellite observations taken of the whole Earth. While plenty of data from current Earth climate monitoring satellites exist, there are limited observations of the unresolved Earth seen as an exoplanet.

Solar sailing provides us with an interesting opportunity to observe the Earth as an exoplanet. For that we can see how the Earth (or the Moon, Venus or Mercury) transits the Sun.  We can further improve the technique of planetary transit spectroscopy by using it to ``find and confirm'' the existence of life on the Earth.

{\bf\em Science objectives}: The pathway to finding biosignatures on habitable worlds depends strongly on the properties of their parent stars. The most common stars in the Milky Way Galaxy are dim, red M-dwarfs. Their habitable zones will be very close to the star, making the systems more accessible for transit observations. Properly interpreting the results of these observations is critical to understanding the formation and history of these planetary systems to see how life-enabling chemicals flow onto worlds ensuring habitability.

Solar sailing propulsion allows us to test, validate and improve the core exoplanetary capability of planetary transit spectroscopy that is now used to find and study atmosphere of exoplanets. The objective here would be to test the technique on a set of known targets that may allow for improvements in the technique, especially for the purposes of unambiguous life confirmation \cite{Green2021}. Transmission spectroscopy of the Earth-Sun system could be used to inform the search for extrasolar life \cite{Mayorga-etal:2021}. Such an approach would allow us to ensure applicability and precision of the methods and tools used to provide evidence of life beyond Earth.

{\bf\em Mission design and requirements}: Observation of Earth at all phase angles (and over different seasons) is a proxy for being able to directly observe an exoplanet separate from its companion star, like what might be accomplished with an advanced coronagraph or a star shade. That requires an orbit around the Earth, and the ability to observe it as an integrated disk. To observe an Earth transit one needs to be over 0.02~AU from the Earth (beyond L2) and able to cross the Earth-sun line, making the disc of the Earth pass in front of the larger disk of the sun (preferably more than once, at different distances). Observing the system with a spectrometer of sufficient accuracy will allow us to see the constituents of the Earth's atmosphere -- and potential habitability and/or life indicators. The key for a sail craft is the ability to actively maintain a 1-year orbit at 1.02--1.06~AU radius, and ``tack'' across the Earth-sun line multiple times to create multiple Earth (and potentially Moon) transits across the Sun.

The mission will be composed of a cubesat class spacecraft, equipped with visible, near-infrared (NIR), and near-ultraviolet spectrometers, as well as a polarimeter with imaging capabilities. The spacecraft will be deployed beyond LEO, reach the needed 0.02~AU distance from the Earth, and will collect data for a minimum of one year to ensure continuous coverage of the Earth from all phase angles and seasons. As the planetary ephemerides are well known, planetary transits are easy to predict for any location in the solar system. A similar approach would allow one to see Venus and Mercury while observing them as exoplanets. For that, a different transiting sail craft could be sent to appropriate region near these planets in addition to Earth.

{\bf\em Instrumentation}: This is a smallsat mission dedicated to taking whole-Earth spectra and polarimetric images in order to measure the temporal variability of observations from an inhabited planet with illumination phase, rotation, cloud cover, and seasons \cite{Coppin2021}. The data would help improve models of habitable planets, guide analysis of exoplanet atmospheric measurements, and inform instrumentation for future missions to search for signs of life on exoplanets. The mission must be able to perform spectroscopic investigations searching for the signs of organic elements in the atmospheres of Venus and Earth, while using the Moon or Mercury as calibration targets. Miniaturized versions of instruments similar to those used by JWST would be suitable for these spectroscopic investigations.

The mission will require a UV/Visible/NIR spectrometer ($\sim$200~nm -- 1.5~$\mu$m) with an imaging polarimeter ($\sim$400--1000~nm) \cite{Coppin2022}. The polarimeter will measure the spectral flux and polarization data of sunlight reflected by Earth. A visible imaging camera with moderate spatial resolution ($\sim$100~km) would aid in interpreting spectral data. This might be the same as the imaging polarimeter, depending on the polarimetry approach. The Star Phoenix concept is well suited for such a program as it uncooled covering the range from $\sim$200~nm$-1~\mu$m, using a MEMS spectrometer.

{\bf\em Technology readiness}:  The trajectory needed for this project does not require reaching velocities beyond what is offered by the current readiness of sailcraft technology. These missions may be flown during Phase I of the overall Sundiver program.

\subsection{Intercepting and probing interstellar objects}
\label{sec:ico}

\begin{figure}
\includegraphics[width=0.9\linewidth]{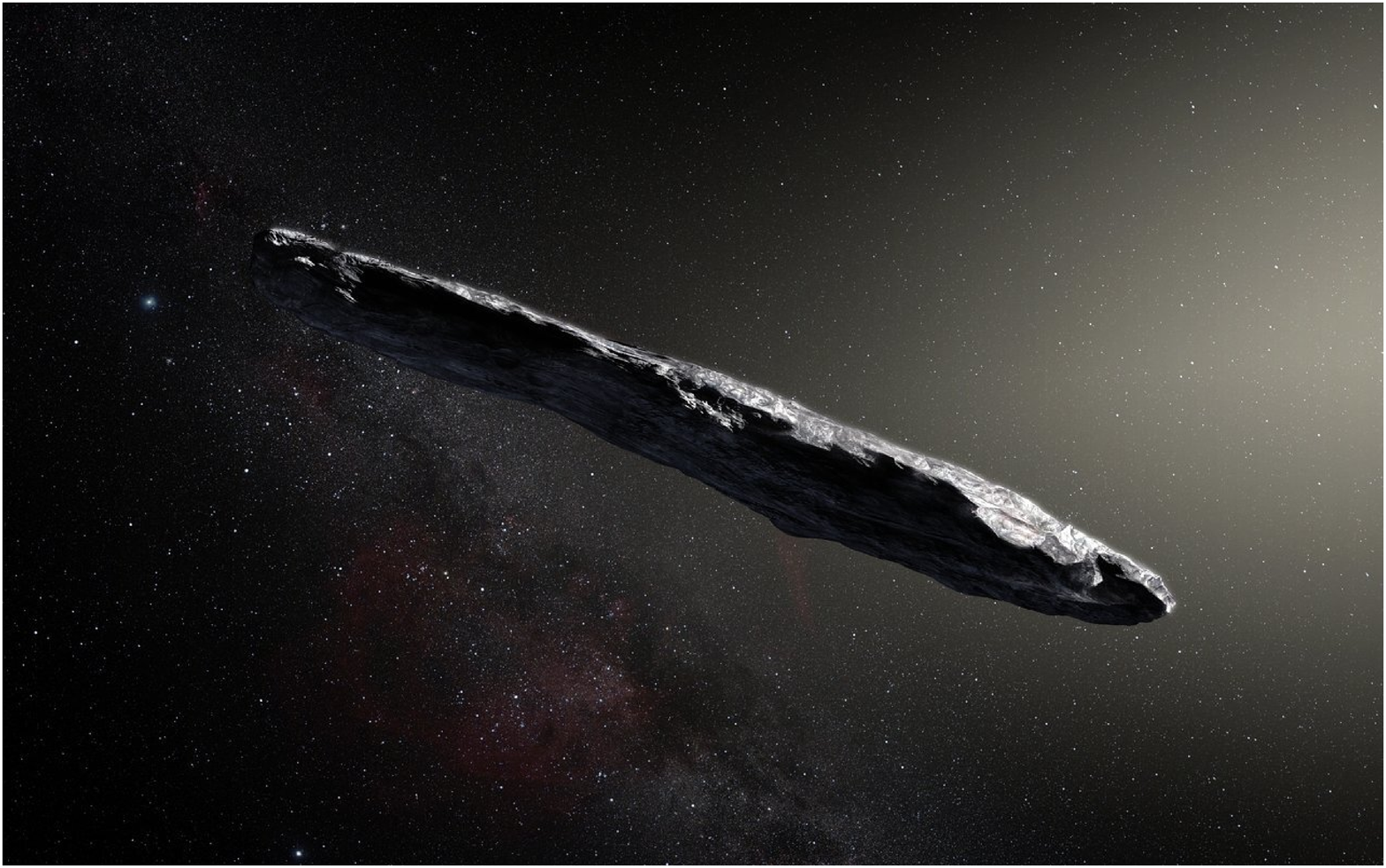}\\
\includegraphics[width=0.9\linewidth]{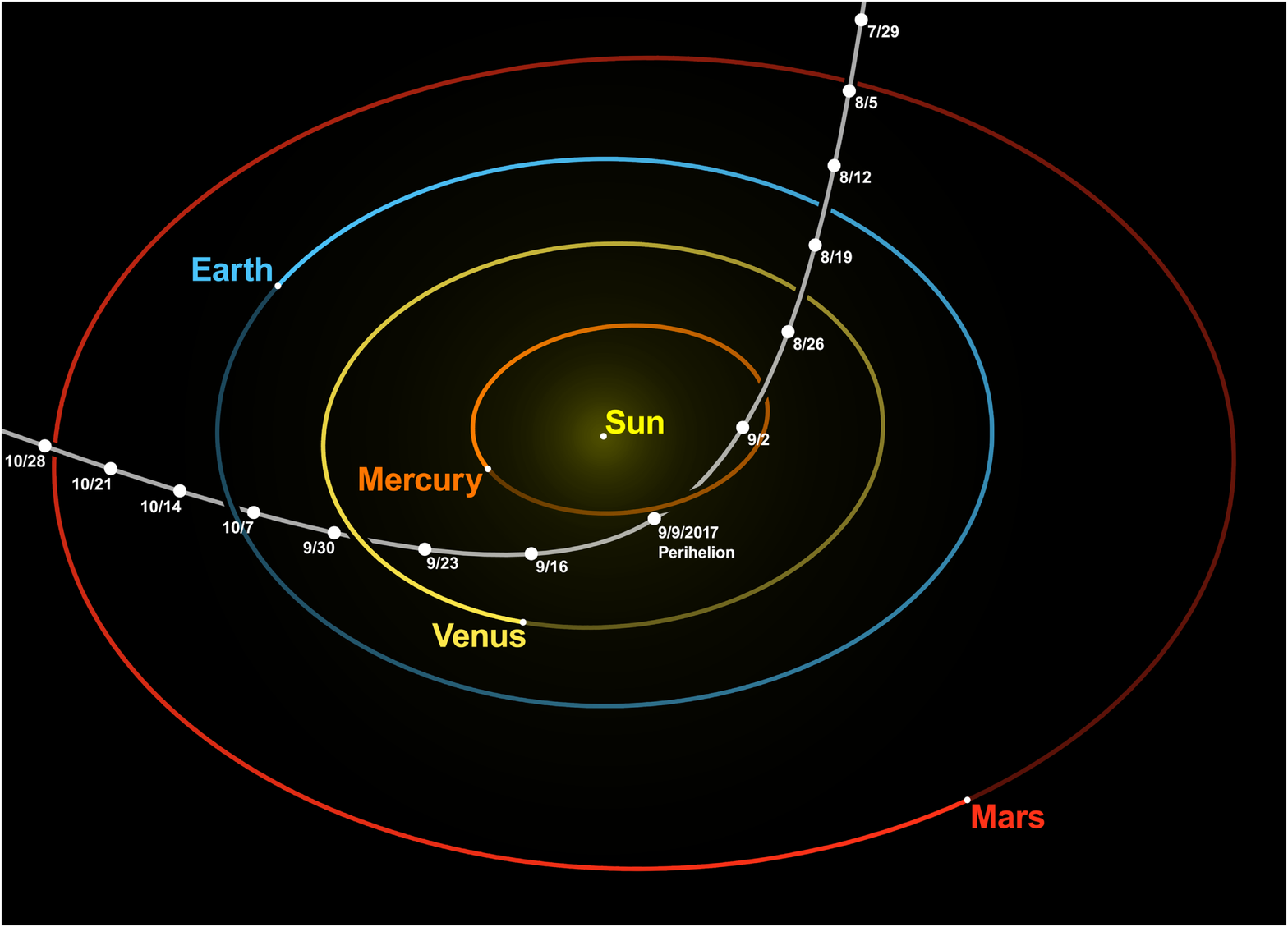}
\caption{\label{fig:oumuamua}Top: Artist's impression of 'Oumuamua (credit:
ESO/M. Kornmesser). Bottom: 'Oumuamua's trajectory in the inner solar system (WikiMedia/Nagualdesign).}
\end{figure}

In October 2017, the first known interstellar object (ISO) to visit our solar system was discovered. The object, named 1I/'Oumuamua (Fig.~\ref{fig:oumuamua}), was detected, tracked and observed as it was moving through the solar system at a heliocentric velocity of $\sim$50~km/s. This discovery allowed for a calibration of the abundance of interstellar objects of its size and an estimation of the subset of objects trapped by the Jupiter-Sun system \cite{Siraj2019}. There should be thousands of 'Oumuamua sized interstellar objects identifiable by Centaur-like orbits at high inclinations, assuming a number density of 'Oumuamua sized interstellar objects of $\sim$$10^{15}~{\rm pc}^{-3}$ \cite{Siraj2019,hoover2022population}. Others estimate a detection of about 15 objects over a 10 year period \cite{hoover2022population}.
In addition, in Aug 2019, 2I/Borisov became the first observed interstellar comet and the second observed ISO after 'Oumuamua. While 2I/Borisov seems to resemble Oort Cloud comets \cite{de2019interstellar}, the nature of 1I/'Oumuamua is still unclear with various proposed hypotheses \cite{bannister2019natural}.

{\bf\em Science objectives}: It is likely that every year, many of these objects pass through the solar system undetected \cite{eubanks2021interstellar}. There may be up $\sim$$10^4$ of objects that originated as ISOs but are now trapped inside the orbit of Neptune \cite{Jewitt2017,Marcos2018}. Imaging or visiting these objects and conducting {\em in situ} explorations would allow us to learn about the conditions in other planetary systems without sending interstellar probes \cite{Hein2022}. Sailcraft on high-energy trajectories also provide a unique opportunity to directly study ISOs transiting through the solar system \cite{Garber2022,miller2022high}. The scientific return from such investigations is invaluable, as comparative studies between an ISO sample return with solar system asteroid and comet sample returns can help us understand the conditions and processes of solar system formation and the nature of the interstellar matter, which is a priority question listed in the Planetary Sciences Decadal Surveys\footnote{\url{https://www.nap.edu/catalog/13117/vision-and-voyages-for-planetary-science-in-the-decade-2013-2022}}$^\text{,}$\footnote{\url{https://nap.nationalacademies.org/catalog/26522/origins-worlds-and-life-a-decadal-strategy-for-planetary-science}}.

Rendezvous with an ISO will help answer unsolved mysteries surrounding these interstellar interlopers \cite{Hein2022}. For example, a rendezvous would allow us to determine the origin of `Oumuamua's nongravitational acceleration: Is it due to cometary outgassing? Due to fractal dust aggregation formed over time? Due to interstellar dust coalescing upon impact? Or is it because of some other exotic origin? The interdisciplinary context of these investigations in a planetary science mission could lead to major progress in several areas of astrophysics. Fractal aggregates, which have been indirectly observed in circumstellar disks, may be the building blocks in protoplanetary disks thus yielding clues for processes guiding formation and evolution of planetary systems in our stellar neighborhood. Sending sailcraft to transient ISOs may allow us to directly access and study the building blocks of exoplanets, which could provide unprecedented constraints on planet formation models.

A close encounter by a swarm of sailcraft with an ISO would help us answer important questions such as: What caused these objects to accelerate? Is there outgassing and if so, what is causing it?  What materials constitute these objects, and how do they compare to typical solar system asteroids or comets? Are ISOs fractal dust aggregates with such a low density that they can be accelerated by radiation pressure?

These questions are fundamental, as the microscopic properties of dust played a key role in particle aggregation during the formation of our solar system. With ISOs being distant messengers from interstellar space, we have an unparalleled opportunity to discover what these transient objects can tell us about our own solar system, its planetary formation, and the interstellar medium to test theories regarding planetary formation. Such discoveries may lead us to reconsider comet and planetary formation models and inform about the properties of the transient interstellar object's parent nebula via its composition.

{\bf\em Mission design and requirements}: Sun-propelled sailcraft are the only means of exploration we have today that could catch up with ISOs moving at tens of km/s and inexpensive enough to be on standby in Earth orbit. If a pursuit had been initiated at the time when 'Oumuamua was discovered, it would only have taken only several months to catch up with it. This capability is within reach today by using solar sailing propulsion and should be considered for ISO exploration. Furthermore, solar sailing will enable flybys, rendezvous, and even sample return missions (using impactors), depending on the trajectory of the ISO and the time left after detection \cite{Hein2022}.

{\bf\em Instrumentation}: A broad range of measurements are sought for ISO observations. They include the characterization of basic physical properties (shape, density, morphology, dynamical properties), compositional properties (elemental composition, mineralogy, isotopes of at least hydrogen, oxygen, nitrogen, and carbon), geophysical/interior properties (porosity, cohesion, magnetic field), geological traits that might inform on origin and possible long-term evolution.

{\bf\em Technology readiness}: The key benefits of Sundivers here is not only the high transit velocity, but also the ability of a solar sailcraft to change orbital inclination. To reach ISOs with sailing such a change is natural. The overall technology readiness for the mission to reach ISOs is high, but some development is needed (sail materials, power sources, precision navigation, autonomous proximity operations, etc.) Such missions may be implemented by the end of Phase II of the Sundiver program.

\subsection{Zodiacal background and interplanetary dust}
\label{sec:zodi}

Missions out of the ecliptic plane and to the outer solar system provide unique opportunities to measure the properties of the local interplanetary dust (IPD) cloud responsible for the zodiacal light (ZL). These clouds are produced by either continuing collision cascades that break up larger bodies, or by evaporation of cometary material. In either case, dust continually fills the interplanetary space. While smaller particles are ejected from the solar system by solar radiation pressure, larger ones may spiral sunward under the action of the Poynting-Robertson effect. As a result, dust populations are continually replenished over time scales of millions of years.

\begin{figure}
\includegraphics[width=0.9\linewidth]{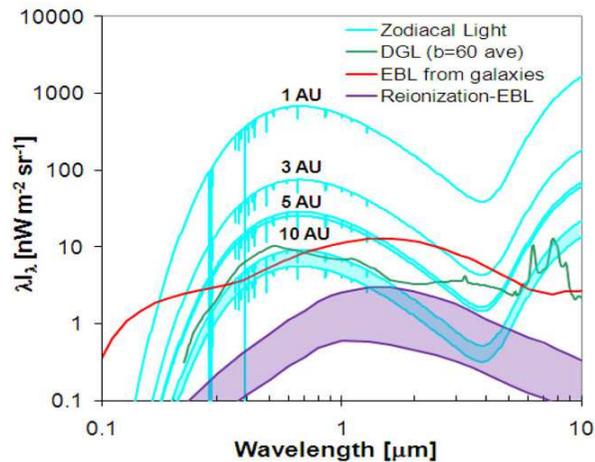}
\caption{\label{fig:zodiac}
Estimated brightness of zodiacal light from 1AU to 10AU based on Pioneer 10 and 11 measurements out to 3AU, with shaded areas indicating the uncertainty in the extrapolation. (From \cite{Bock-etal:2012}.)}
\end{figure}

The main sources of zodiacal dust in the inner solar system and likely Kuiper belt dust as well are collisions of bodies in the main asteroid belt, lying between the orbits of Mars and Jupiter, and the Kuiper belt, lying beyond the orbit of Neptune, together with cometary contributions. Dust belts produced by similar processes are also commonly found around nearby stars, tracing planetesimals in these systems just as they do in our solar system. A study of the IPD is of broader scientific interest because of its connection with extrasolar planetary systems; it could tell whether our own planetary system is an outlier in some important way.

{\bf\em Science objectives}: Missions traveling toward Saturn's orbit and beyond can provide information on morphology and variation in composition and temperature as a function of distance from the sun in order to better understand the processes shaping the local IPD cloud and help interpret measurements of dust in exoplanetary systems. Solar sailing missions traveling out of the ecliptic plane can provide information on the scale height of the IPD and three dimensional variation in composition and temperature\footnote{Note that similar investigations may be conducted to explore the far-UV background \cite{Kulkarni:2022}.}.

{\bf\em Mission design and requirements}:
Missions to characterize the zodiacal light and IPD can have any trajectory although most interesting would be missions traveling radially outward past 5 AU and/or missions out of the ecliptic plane. The intensity of the zodiacal light is expected to fall roughly according to the radius from the sun squared along the ecliptic plane and by roughly an order of magnitude at a distance of 0.5 AU perpendicular to the ecliptic plane at a radius of 1 AU from the sun \cite{Leinert-etal:1998}. However, contributions  to the IPD from cometary tidal tails would be expected to have a more spherical distribution \cite{Carleton-etal:2022}.

{\bf\em Instrumentation}:
The instrumentation required to characterize the IPD and zodiacal light includes optical and near-infrared photometers and spectrometers as well as dedicated dust impact detectors for in situ measurements such as the measurements of dust impacts on the Juno spacecraft star camera and solar arrays during solar system transit \cite{Jorgensen-etal:2021}.
Distribution of interplanetary dust detected by the Juno spacecraft and its contribution to the Zodiacal Light \cite{Benn-etal:2017,Jorgensen-etal:2021}. An instrument flying toward the outer solar system could measure the radial distribution of the IPD and map resonant enhancements and band structures in the zodiacal dust influenced by planetary bodies. It could study the compositional distribution of dust and determine if it arises from comets, asteroids, or both, from the inner to the outer solar system. This instrumentation would provide important data on how the dust diffuses outward to fill the solar system; how dust properties (i.e., density, composition, size, etc.) change as a function of radial position; and how the observed dust populations compare with those seen in exoplanetary systems. Such important science investigations may be easily performed even with first-generation sailcraft.

{\bf\em Technology readiness}: New thermally resistant reflective materials, thermally protective sailcraft design and avionics and autonomy are needed for the close perihelion required for high transit velocities. On-board power and communications downlink capacity are the main capabilities that must be developed. Missions addressing these objectives are suitable for Phase II of Sundivers.

\subsection{Cosmic background and the reionization epoch}
\label{sec:cb-reionize}

Measurements of the absolute intensity of the extragalactic background light (EBL) in the optical and near-infrared constrain the integrated star formation history of the universe \cite{Hauser-Dwek:2001,Cooray:2016}. The ability to accurately measure the EBL depends strongly on the ability to remove foreground emission including scattered light from the Earth's atmosphere or sky glow, scattered light from instrument optics, diffuse galactic light, and scattered sunlight from interplanetary dust in the solar system or the zodiacal light. Measurements of the EBL from deep space reduce or eliminate two of the main foregrounds: sky glow and zodiacal light. While there has not been a deep space mission dedicated to the measurement of the EBL, data from science and navigation instruments operating during the cruise phases on board planetary science missions including Pioneers 10 and 11  \cite{Toller:1983,Matsuoka-etal:2011,Matsumoto-etal:2018} and New Horizons \cite{Zemcov-etal:2017,Lauer-etal:2021,Lauer-etal:2022}  have been used to constrain the EBL and give values for the EBL that are a factor of two brighter than the integrated light from galaxies as measured by ground-based and near-Earth observatories \cite{Windhorst-etal:2022}.

To avoid contamination by zodiacal light, there is a need to perform measurements of the cosmic background outside the zodiacal dust cloud, beyond the orbit of Saturn. Conducting EBL observations with sufficient resolution to identify individual stars makes it possible to strongly suppress these foregrounds.

{\bf\em Science objectives}:
Access to the outer solar system allows for a significant reduction in zodiacal brightness, enabling precise measurements of the extragalactic back- ground light (EBL) \cite{Lauer2021}. It also allows for a deep search for redshifted Lyman alpha photons from reionization, which is critical to interpret cosmological data (Fig.~\ref{fig:zodiac}).  From vantage points deep in the outer solar system and beyond, sailcraft can study fundamental questions in astrophysics and cosmology. By measuring the EBL, they will be able to address questions such as how the Universe originated and evolved to create its galaxies, stars, and planets that we see today. The EBL is a cornerstone measurement needed to probe the fossil record of star formation and galaxy assembly from the first stars to the present day.

{\bf\em Mission design and requirements}:
To allow for high-quality EBL data, the mission must be in the outer parts of the solar system beyond 5 AU or out of the ecliptic plane; no precision pointing is required. The higher the velocity, the faster Sundivers will reach the desired region.

{\bf\em Instrumentation}: In order to make precise measurements of the EBL at optical and near-IR wavelengths, especially at those corresponding to the redshifted Lyman alpha emission from reionization, spectroscopic imaging is required. A spectral resolution of $R\gtrsim100$ at wavelengths from 0.8--1.5~$\mu$m is achievable with an instrument similar to the BIRCHES IR spectrometer flown on the Lunar Ice Cube mission \cite{Clark-etal:2019}. Ground-based imaging data at arcsecond resolution can be used to remove the galactic and zodiacal foregrounds \cite{Primack2008}.

{\bf\em Technology readiness}:
New thermally resistant reflective materials, thermally protective sailcraft design and avionics and autonomy are needed for the close perihelion required for high transit velocities. On-board power and communications downlink capacity are the main capabilities that must be developed. Missions addressing these objectives are suitable for Phase II of Sundivers.

\subsection{Testing gravity on the way out of the solar system}
\label{sec:test_gravity}

Sailcraft on fast hyperbolic trajectories may be used to test the foundations of relativistic gravitation on scales never attempted. Precision navigation of spacecraft from 1--100~AU enables very powerful tests of long-range modifications of gravity (similar to the investigations of the Pioneer anomaly \cite{Anderson-etal:2002,Nieto-Turyshev:2004,Turyshev-Toth:2010,Turyshev:2012}). These measurements can help refine constraints on the range and coupling constant of Yukawa-type extensions of Newtonian gravity: effects due to such forces increase with distance, leading to a possible improvement by a factor of 100 compared to current results \cite{Turyshev2008}. Such an experiment can also help improve limits on the post-Newtonian parameters of generalized theories of gravitation, in particular Eddington's $\gamma$-parameter, which measures the effects of spatial curvature. Limits on the minimum acceleration $a_0$ and constraints on the interpolating function  ($g/a_0$) of Modified Newtonian Dynamics (MOND) \cite{Blanchet2011,Milgrom2014} can be refined, along with possible violations of the weak equivalence principle.

MOND predicts a phenomenological modification of Newton's second law when the (gravitational) acceleration becomes less than the MOND parameter, $a_0 = (1.22\pm 0.33) \times 10^{-10}~{\rm m}{\rm s}^{-2}$. The ``MOND radius'' of the Sun is huge---$7\times 10^3$~AU---and even out there, although MOND should dominate (in  theory) it would not be easy to test. However, there are ``saddle points'' where the sum of the accelerations of the Earth, Moon, and Sun cancel. There might be large non-Newtonian effects right at the saddle point \cite{Bekenstein2006,Magueijo2012,Banik2019}.

Also, dark matter may influence the trajectories of spacecraft sufficiently far from the Sun
\cite{Belbruno-Green:2022}, thus providing conditions necessary for its detection and study.

{\bf\em Science objectives}: An experiment to precisely measure gravity at large distances from the Sun would establish stringent limits on possible violations of Newtonian gravity while probing for the presence of dark matter in the outer solar system. It will be important for the validation of various theories of modified or massive gravity theories that are proposed as alternatives to dark matter or the cosmological constant. Experimental confirmation of new fundamental forces would provide an important insight into the physics beyond the Standard Model. These results will help bridge the presently extant gap between the data in the solar system and astronomical data obtained on scales that are several orders of magnitude larger, characterizing the dynamical gravitational behavior of star clusters and galaxies.

\begin{figure*}[t!]
\includegraphics[width=0.75\linewidth]{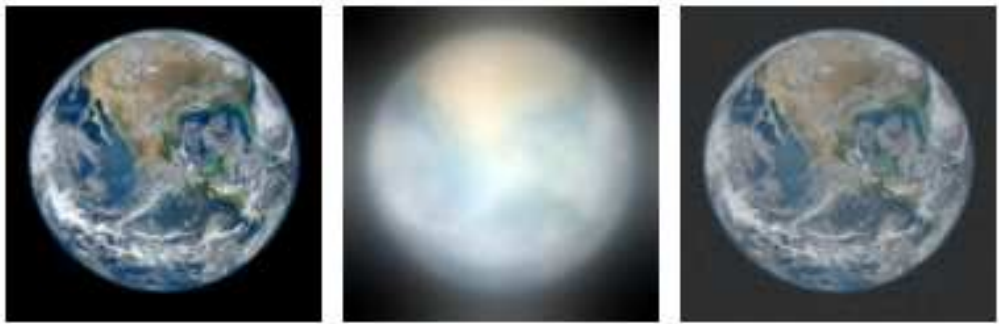}
\caption{\label{fig:SGLsimul}Imaging simulation with the SGL. Left: original RGB color image of an Earth-like exoplanet with $1024\times 1024$ pixels; center: image blurred by the SGL, sampled at an SNR of $\sim$$10^3$ per color channel, or overall SNR of $3\times 10^3$; right: image deconvolution result (see \cite{Turyshev2020,Turyshev2022} for details).}
\end{figure*}

{\bf\em Mission design and requirements}: The biggest difficulty of a saddle point mission is that these saddle points are small and non-inertial. This makes flythrough times short, limiting the measurement signal-to-noise ratio. Two or three sailcraft would need to make repeated passes through or near the saddle point (say, one per month) and act as a gradiometer. However, if a fast sailcraft could stay in or near a saddle point for extended periods, hours rather than seconds, a single spacecraft might suffice. The mission would require good tracking, a good gradiometer (e.g., the MEMS-based accelerometer from Glasgow University\footnote{\url{https://www.gla.ac.uk/research/beacons/nanoquantum/wee-gglasgowsgravimeter/}}), and a good clock.

{\bf\em Instrumentation}: To test the gravitational inverse square law with Sundivers we need to achieve a precision navigation of these microsats at the acceleration level below $10^{-10}$~m/s$^2$. This level of acceleration knowledge can be obtained using modern-day RF communication systems that are used for radio science investigations. Sundivers may require miniaturization of these systems.

{\bf\em Technology readiness}: Some technology development is needed (sail materials, power sources, precision navigation, autonomous proximity operations, etc.) These missions may be implemented early during late Phase II or in Phase III of the Sundiver program.

\subsection{Exoplanet imaging: the solar gravitational lens}
\label{sec:sgl}

Most of the work described in this paper is an outgrowth from the NIAC Phase I--III studies, led by one of us (SGT), on ``Direct Multipixel Imaging and Spectroscopy of an Exoplanet with a Solar Gravitational Lens Mission'' (2017--2022) \cite{Turyshev-etal:2020-PhaseII,Turyshev-etal:2020-PhaseIII}.  A mission to the SGL's focal region is enabled by interplanetary smallsats, which was the reason for developing the TDM to demonstrate the technologies (see Sec.~\ref{sec:tdm}). Such a mission would be the ultimate fast interplanetary smallsat, and many such missions conducted at low-cost would enable remote exoplanet exploration.

Imaging extrasolar terrestrial planets combined with spectroscopy is probably the single greatest remote sensing result that we can contemplate in terms of galvanizing public interest and support for deep-space exploration. However, direct multipixel imaging of exoplanets requires significant light amplification and high angular resolution. With classical optical instruments, we face a sobering reality. To capture even a single-pixel image of an Earth-like exoplanet at a distance of $z_0=100$~light years at the near-IR wavelength $\lambda=1~\mu$m, a $d=1.22(\lambda/2R_\oplus)z_0 \sim 90.5$~km telescope aperture or baseline would be needed. An interferometer network, even if it involves thousands of $\sim$30~m telescopes, would require integration times of $\sim$$10^5$~years to achieve ${\rm SNR}\gtrsim 1$ against the exo-zodiacal background. Resolved multipixel imaging would necessitate even larger (thousands of km) baselines. These scenarios are impractical, giving us no hope to spatially resolve and characterize exolife features.

To overcome these challenges, we consider the solar gravitational lens (SGL) to produce high-resolution, multipixel images of exoplanets \cite{Turyshev2017,Turyshev2020,Turyshev2022,Turyshev-Toth:2023}. This lensing results from the diffraction of light by the solar gravitational field, which focuses incident light at distances $>$$548$~AU behind the sun. The properties of the SGL are quite remarkable, including light amplification of $\sim$$10^{11}$ and angular resolution of $\sim$$10^{-10}$~arcsec \cite{Turyshev2022}. The SGL is our only means to obtain spatially resolved spectroscopic data that could show the presence of current life and obtain ``close-up'' images of an exo-Earth\footnote{For an overview of the imaging approach, please see \url{https://www.youtube.com/watch?v=NQFqDKRAROI}} (see Fig.~\ref{fig:SGLsimul}).

{\bf\em Science objectives}: A meter-class optical telescope with a modest coronagraph operating in SGL's focal region, beyond 650 AU, can yield a $(250\times 250)$-pixel image of an Earth-like exoplanet even at 30~pc in just 12 months, which is not possible by other known means. Even in the presence of the solar corona, the signal is strong enough to image of such an object with $\sim$50~km scale resolution of its surface, enough to see signs of habitability, observe seasonal changes, image surface topography, obtain spectroscopy of the atmosphere and model the climate. Closer targets will yield higher-resolution images in shorter times \cite{Turyshev2022}.  In addition, the SGL has unique capabilities for a spatially-resolved spectroscopy in the near-IR band within 2--20 $\mu$m \cite{Turyshev2022b}.

{\bf\em Mission design and requirements}: Turyshev {\em et al.} \cite{Turyshev-etal:2020-PhaseII} have developed a new mission concept to deliver optical telescopes to the SGL's focal region and then to fly along the focal line to produce high resolution, multispectral images of a potentially habitable exoplanet. The proposed multi-smallsat architecture \cite{Helvajian2022} uses solar sailing and is designed to perform observations of multiple planets in a target extrasolar planetary system\footnote{For an overview of solar sailing, please see \url{https://www.youtube.com/watch?v=o6bysOWOK6U&t=868s}} . It allows reduced integration time and accounts for the target's temporal variability, which helps to ``remove the cloud cover'' and also to explore the spectroscopic contents of the planetary atmosphere.

{\bf\em Instrumentation}: The prospect of getting an image of an exoplanet and to spectroscopically detect and characterize life there is compelling. With the SGL, we can search for possible biosignatures on planets where life like ours would have emerged and modified the atmosphere in a way that can be detected by remote sensing, i.e., spectroscopic observation of O${}_2$, O${}_3$, H${}_2$O, CO${}_2$, CH${}_4$ gases \cite{Turyshev2022b}. If a solar sailing mission to the SGL focal region can provide spectroscopic proof of life on an exoplanet, it would qualify as one of the most exciting advances of scientific discovery in history. The main instrument needed for this mission is a coronagraph-equipped meter-class telescope operating either in visible of near-IR.

{\bf\em Technology readiness}: As sailcraft technology matures (see Sec.~\ref{sec:sdconcept}),  the  Sundiver concept may open a path for humanity to prepare for interstellar missions to nearby stars. Clearly, this is not yet for tomorrow, nor the day after, as we must learn to walk before running. One of the missions on that path may be a mission to image an inhabited exoplanet in our stellar neighborhood using the SGL.

This mission is the ultimate objective of the Sundiver program, requiring the highest transit velocity combined with precise navigation and high levels of autonomy. Some technology development is needed (sail materials, power sources, precision navigation, autonomous proximity operations, in-flight aggregation, autonomous operations in the deep space, etc.) This mission will be implemented in Phase III of the overall Sundiver program, providing a great motivation for the new paradigm.

\section{From HEO to cislunar and beyond}
\label{sec:moon}

Low-thrust propulsion technologies such as solar sails and electric propulsion are key to enabling many space missions which would be impractical with chemical propulsion. Taking advantage of their small size and shorter development timelines, small spacecraft are increasingly capable as both rapid precursor missions and as components of a cost-effective in-space infrastructure. In that respect, the ongoing developments of the TDM for the advanced sailcraft with high $A/m$ ratio (see Sec.~\ref{sec:paradigm}) offer exciting new opportunities. The new sailcraft design with articulable vanes allows for missions with high dynamic $\Delta v$ all the way from high LEO to the Moon and beyond. Benefiting from the renewed emphasis in lunar exploration, these missions will help to evolve the Sundiver sailcraft  to full technical maturity while using it for the development of the infrastructure in cislunar space.

For example, while still in the Earth's proximity, either at MEO\footnote{Medium Earth orbit (MEO): \url{https://en.wikipedia.org/wiki/Medium_Earth_orbit}} or HEO\footnote{High Earth orbit (HEO): \url{https://en.wikipedia.org/wiki/High_Earth_orbit}}, modern sailcraft can be used for the following important applications:
{}
\begin{itemize}
\item Assessment and removing of LEO debris. It is known that there are at least 500 tons of debris, consisting mostly of discarded boosters and components as well as inoperative spacecraft, are present below 1100 km of altitude, threatening LEO satellites with collisions\footnote{\url{https://orbitaldebris.jsc.nasa.gov/quarterly-news/pdfs/odqnv15i2.pdf}}.
A collision between any two of them would double the number of objects in the LEO catalog. Thus, understanding of the stability, rotation rates, structural integrity, micrometeoroid impacts and UV effects on the debris would provide critical information to improve the models and long-term threat assessments. The same sailcraft maybe used to de-orbit larger pieces of debris, yielding viable commercial business models.
\item Rendezvous and proximity operations (RPO): With their ability to match the orbit of objects in MEO, GEO\footnote{Geostationary Earth orbit (GEO): \url{https://en.wikipedia.org/wiki/Geostationary_orbit}}, and cislunar space, sailcraft could be used to rendezvous with  various assets on these orbits for the purposes of delivery and/or inspection. They can also serve as calibration sources for activation of RF payloads\footnote{\url{https://www.esa.int/Enabling_Support/Space_Engineering_Technology/RF_Payloads_Technology}}. These are standard secondary payloads for all MEO and GEO launches to support instrument activation and deployment.
\item Modern sailcraft may serve as cost-effective means for generating large constellations for relay, collection, inspection, and monitoring missions.
\end{itemize}

Closer to the Moon, technology development may focus on three technical areas needed for lunar-bound missions:
\begin{inparaenum}[i)]
\item Use of small spacecraft to help provide lunar communications and navigation services,
\item Small spacecraft propulsion for lunar missions and potential return of lunar samples using small spacecraft, and
\item Small spacecraft electrical power and thermal management systems tailored for the distant and harsh environment between the Earth and the Moon.
\end{inparaenum}
These objectives may lead to several projects aiming at the development cislunar infrastructure, including those related to space weather monitoring, navigation, observations, power generation, and communication:
{}
\begin{itemize}
\item Space weather monitoring. Advanced solar sailing vehicles may provide an early warning system operating on a nearly continuous and long-term basis while operating on various orbits -- from HEO to cislunar and to Earth-trailing. In fact, modern sailcraft could enable missions to observe the solar environment from unique vantage points of interest to heliophysics \cite{Liewer-etal:2013,Vourlidas-2015}, including maintaining a satellite in a position sunward of Lagrange point L1\footnote{\url{https://en.wikipedia.org/wiki/Lagrange_point}}. In fact, solar sails are very valuable at all other  Language points from L1 to L5.

\item Solar sailcraft may be used to enable cislunar navigation which can be done in the manner similar to modern global navigation satellite systems (GNSS, e.g., GPS, Galileo, GLONASS, BeiDou, etc.) For that purpose the sailcraft with RF communication relay capabilities could establish a set of artificial Lagrange points (ALP) in the Sun--Earth+Moon system \cite{Pan-etal:2017} with the lines of sight to areas of interest on the lunar surface. Compared with the classical Lagrange points, these ALPs would benefit from the sailcraft ability to reach dynamical equilibria and out-of-plane placement that would allow for the construction of a disperse sailcraft constellation. Although under the influence of lunar gravity solar sails cannot be stationary at the ALPs nor they can move in bounded orbits nearby, the use of articulable vanes enables the required control mechanism.
\item Modern space missions are increasingly venturing across cislunar space, requiring expansion of space awareness functions \cite{Ewart-etal:2022}, given importance for a ``pole-sitter'' mission architecture.  A key advantage of a pole-sitter is that it has a position well outside the ecliptic plane and offers a unique, in some cases orthogonal viewing geometry that has to be explored for operational deployment. If implemented with traditional means, propulsion would be a challenge for a pole-sitter satellite. On the other hand, a combination of an advanced solar sail architecture and solar electric propulsion (SEP)\footnote{As was done on the JAXA's ICARUS mission and is planned for OKEANOS mission \url{https://en.wikipedia.org/wiki/OKEANOS}.} may be used to accomplish such an objective.
\item With their large surface area, the solar sailing smallsats may be used to collect solar energy in space (e.g., by relying on the photovoltaic elements embedded on the sail, as discussed in Sec.~\ref{sec:paradigm}) and to beam it down to the lunar surface via laser links, which is especially important during lunar nights.
\item To conduct exploration activities on the far side of the Moon, communication is important.  Solar sails may be used to enable relay communication between the assets on the surface of the Moon and terrestrial ground-based facilities. These advanced sailcraft systems could be used to construct sparse aperture arrays. These are a cost-effective means for generating large constellations for relay, collection, and monitoring missions. Also, by articulating the vanes, the vehicle can compensate for pointing errors in the beam and remain on course.
\end{itemize}

Missions, addressing the objectives above, can be implemented early on in Phase I of the Sundiver program (see Sec.~\ref{sec:sdconcept}). As a result, with their ability to move on a recurring and continuous basis between the HEO to cislunar space and beyond, the inexpensive solar sailcraft with advanced performance may provide us with the needed delivery, inspection, and exploration capabilities. Relying on the abundant solar radiation pressure to move in their paths, using embedded PV elements  to have access to solar power to charge their batteries, and employing phased-arrays for efficient communication with both terrestrial and/or lunar facilities, these vehicles are uniquely suited for many missions. As such, these TDM-derived advanced sailcraft  are ready to play a prominent role in the ongoing exploration of the cislunar/translunar space and beyond, thus contributing to the technical maturity of the Sundivers and their ultimate use in the deep space.

\section{Conclusions}
\label{sec:concl}

The year 2022 marked the 50th anniversary of the launch of Pioneer 10. Since then, only five other spacecraft have made it past Jupiter's orbit (Pioneer 11, Voyager 1 and 2, Cassini--Huygens and New Horizons) and no new ones are planned for at least another decade. Missions to the outer solar system take decades to develop and carry out, and this is reflected in their cost: typically, several billion dollars. Driven by such cost options (augmented by the requirement for large launch vehicles), these missions are designed using a highly risk-averse philosophy, slowing down development and further increasing the price.

Long design and construction lead times have an additional consequence: the technology incorporated in such spacecraft is often dated, reducing the quality of the science conducted. This is compounded by the time it takes to reach destinations in very deep space as travel times to the outer solar system often exceed a decade, with speeds constrained by the fundamental limitations of chemical propulsion.

None of this diminishes the accomplishments of those big missions like Galileo and Cassini--Huygens, and the planned Europa Clipper.  Nor does it contradict the premium the science community places on future such missions, as evident in National Academy of Sciences decadal studies for missions to the Saturn system and later to Uranus.  If realized, they will be brilliant.  It only emphasizes the paucity of missions that currently can be approved to fly to the far outer solar system.  There is a way, we suggest, to develop something that can fill in the gaps between big missions and their inevitable delays.

Using the current approach, designing and building a spacecraft to travel the billion miles to Uranus takes over a decade. The cruise time to the solar system's third-largest planet adds an additional 15 years. A young student who sees such a project being initiated will reach the age of 50 before seeing any science returns. It does not have to be this way.

A rapidly maturing technology paradigm offers much faster access to the outer solar system at substantially lower cost: the Sundiver paradigm. Small spacecraft, utilizing solar sails, passing close to the Sun can achieve much greater velocities, reaching targets in the outer solar system significantly faster and cheaper. They can also be launched with small launch vehicles or even as rideshares. The new paradigm for solar system exploration provides the science community with relatively inexpensive and frequent access to distant regions of the solar system, defying $\Delta v$ limits and obtaining $4\pi$ observational coverage around the Sun. It permits missions with faster revisit times and more visits using the latest technology upgrades, at significantly lower overall cost. The concept is based on the idea that small and minimally capable sailcraft can be placed on fast solar system transit trajectories. Additional capabilities may be unlocked with in-space assembly to form more mission-capable satellites.

Today, solar sails can reach solar system transit velocities of $\sim$5--10~AU/yr (more than twice that of New Horizons), cutting travel times within the Solar System by half. Such spacecraft could reach Jupiter in a year and reach Saturn's moons in less than three, far outpacing past missions. Crucially, the cost of a lightweight, agile Sundiver mission may be in the range of \$30--75M, which compares very favorably to the \$2--5B cost of a typical flagship-type deep space mission.

Although Sundivers would not be able to collect as much scientific data as a flagship mission, each of these missions offers unique and clearly defined scientific value. Moreover, the low-cost, low-risk approach of the Sundiver program allows planetary reconnaissance missions that would be several times faster and almost a hundred times less expensive than the current approach. These missions will provide major opportunities for miniaturization of key scientific sensors, engaging a wide range of industrial partners.

While this paper addressed exploration activities in the deep solar system, other opportunities exist, especially in HEO and in cislunar space. With abundant solar energy, solar sailing vehicles placed on low-thrust trajectories could travel between the Earth and the Moon or stay in cislunar space indefinitely while being involved in exploration-related efforts and services. Photovoltaic elements embedded in the sail could provide electric power, while some parts of the sail could be reshaped to enable high bandwidth RF communications. The result is an autonomous system with many practical uses in support of ongoing lunar exploration efforts.

There are several key technologies that make Sundiver missions possible, including heat-resistant materials and structures; agile, capable small (20--50\,kg) spacecraft; modular power systems; miniaturized sensors; and sophisticated onboard computing for greater in-flight autonomy. Developing these technologies into viable missions requires a more agile, risk-tolerant approach than the approach used by NASA for flagship missions. Such high-risk, high-reward research and development is well suited for private industry working in partnership with the appropriate US Government agencies, including NASA, NSF, the DOD, and the DOE. With proper coordination among players, science could benefit tremendously from these developments in the coming decade. Sundivers may provide a bold vision for renewed planetary and heliospheric exploration, leveraging the strengths and risk tolerance offered by private enterprise.

The Sundiver concept allows access to distant regions of the solar system by spacecraft that could have an exhaustive list of instrumentation. It also permits the release of instrumentation during a flyby. This is possible by enabling instrument assembly from a set of modular components, all individually delivered by sailcraft, accelerated to high velocity, and then physically aggregated to form a larger and more capable spacecraft which continues at high velocity. This concept reduces cost by modularizing systems and invoking mass production technologies, permits repurposing of systems and increases the number of possible missions in a disaggregated approach.

Miniaturized intelligent space systems offer major advantages for science investigations. We expect that objects of major scientific significance, like debris from extrasolar planets and pristine building blocks of our solar system, as well as the possibility to image exoplanets in our interstellar neighborhood, will foster novel approaches to space exploration. This vision may be achieved with standalone science missions or through partnerships with industry and philanthropic organizations that make use of the increasing number of small satellites. Private companies with internal and government support are paving the way for large-scale manufacturing of capable space platforms at low recurring costs and offer a business model for future endeavors. With proper coordination among key players, science could benefit from these developments in the coming decade, providing a bold vision for renewed planetary exploration.

Small sailcraft can be duplicated at low cost and individual payloads can be tailored to individual targets.  Ideally, this creates a program starting with the TDM described in Sec.~\ref{sec:tdm}.  It could be ready for flight as early as 2025. The first science mission in terms of technical readiness would be the solar polar orbiter, as it does not demand any more technology readiness than the demonstration mission.  Fast solar system missions would follow, going closer to the Sun, achieving higher hyperbolic velocities and using technology development for bigger sails and lighter components. As noted, none of these very small and special purpose missions would replace the desire and planning for the larger, more expensive flagship class missions for solar system science.  But they would provide us with initial reconnaissance data while we wait for the orbiters and landers, helping to improve the science return from our solar system exploration efforts.

This exciting new paradigm for deep space exploration is nascent, waiting for the right set of stakeholders to recognize its value. Smart public-private partnerships have proven their ability to allocate risks and rewards, supercharging innovation and efficiency. In such a realm, clearly delineated responsibilities, and a variety of outcomes-based compensation instruments might produce a capability that transforms our relationship with interplanetary space forever. It is the time for us to embark on such an exciting path.

Ultimately, and perhaps most importantly considering the fundamental {\it raison d'etre} for a government-funded space program, the development of missions described herein will lead to readiness for a mission to the solar gravitational lens (SGL) focal region, offering the opportunity to actually see life on other worlds, habitable (or perhaps even inhabited) planets in other star systems.

\begin{acknowledgments}
We would like to express our gratitude to our many colleagues who have either collaborated with us on this manuscript or given us their wisdom. We specifically thank Harry Atwater, Linden Bolisay, Penelope Boston, Robin Canup, Darrel Conway, Bethany Ehlmann, Juan M. Fernandez, John Hanson, Les Johnson, Sarah Johnson, Shri Kulkarni, Avy Loeb, Philip Lubin, Gregory L. Matloff, Francis Nimmo, Merav Opher, Greg Pass, Mason Peck, Elaine Petro, Scott Schick, Thomas Svitek, Grover Swartzlander, and Edward Witten who provided us with valuable comments, encouragement, and stimulating discussions of the various topics discussed in  this document while it was in preparation.

The Aerospace Corporation provided valuable input on many aspects presented in this paper and also conducted an independent cost estimate for the TDM. Our thanks go to Thomas Heinsheimer, Henry Helvajian, and John P. McVey for their interest, encouragement, and comments on various points raised in this document.

We are grateful to our many European colleagues. In particular, our gratitude goes to Matteo Ceriotti, Bernd Dachwald, Lamberto Dell'Elce, Benjamin Fernando, J. Thimo Grundmann, Alesia Herasimenka, Vaios J. Lappas, Colin R. McInnes, and Giovanni Vulpetti who benefited us with their insightful comments and suggestions.

Breakthrough Initiatives supported several  meetings with the science community for a broader discussion of potential Sundiver missions. We also thank Tom Kalil of the Schmidt Futures for  encouragement and valuable advice on various programmatic points discussed here.

We thank our colleagues at JPL for their encouragement, support, and advice regarding this manuscript.  We especially appreciate interest and valuable feedback from David A. Bearden, Julie Castillo-Rogez, Anthony Freeman, Lorraine Fesq, Paul F. Goldsmith, Keith Grogan, Damon Landau, Gregory Lantoine, Paulett C. Liewer, Rosaly Lopes, Charles Norton, Humphrey W. Price, Robert Staehle,  Richard Terrile, and Neal J. Turner  who have kindly provided us with  insightful comments and valuable suggestions on various aspects of the manuscript.

Part of this work was funded by the NASA Innovative Advanced Concepts (NIAC) Program through the 2020 NIAC Phase III grant on ``Direct Multipixel Imaging and Spectroscopy of an Exoplanet with a Solar Gravitational Lens Mission'' (with S.G. Turyshev being the PI).  Our gratitude goes to Jason E. Derleth, Michael R. Lapointe, John C. Nelson,  Katherine M. Reilly, Frank Spellman, and Ronald E. Turner of NIAC for their support, interest, and encouragement.

Many of our meetings were held at the Keck Institute for Space Studies (KISS) at Caltech. We thank Thomas A. Prince and Michele A. Judd of the KISS for their hospitality, encouragement, and support.

The work described here, in part, was carried out at the Jet Propulsion Laboratory, California Institute of Technology, under a contract with the National Aeronautics and Space Administration (NASA) (80NM0018D0004). VTT acknowledges the generous support of Plamen Vasilev and other Patreon patrons.
Pre-decisional information -- for planning and discussion purposes only. The cost information contained in this document is of a budgetary and planning nature and is intended for informational purposes only. It does not constitute a commitment on the part of JPL and/or Caltech.
\end{acknowledgments}

\appendix

\section{Representative technology readiness}
\label{app:techready}

The Sundiver bus system shall provide the required support to the payload needed to fulfill the mission and science requirements, including:
\begin{inparaenum}[1)]
\item Structural and mechanical support;
\item Thermal control,
\item Electrical power,
\item Attitude and orbit control,
\item Onboard computing,
\item Onboard system management,
\item Telecommunication.
\end{inparaenum}

Table~\ref{tb:tech} provides an overview of the current (c.2023) technology readiness of these subsystems.

\begin{table*}[!hbt]
\caption{\label{tb:tech}High-level overview of the Sundiver-specific technological maturity of various subsystems.}
\begin{tabular}{|p{3.5cm}|l|c|l|}\hline
{\bf Subsystem}&{\bf Potential solutions}&{\bf TRL}&{Remarks}\\\hline\hline
Structural&Truss primary structure&7/8&$\sbullet$ ISS heritage\\
&&&$\sbullet$ May require in orbit assembly\\\hline
Mechanisms \& actuators&Brushed DC motors \& controllers&8&$\sbullet$ Main use is vane (solar sail) actuation\\
&&&$\sbullet$ Controllers to be qualified for deep
space\\
&&&$\sbullet$ Suited for infrequent use on long lifetime missions\\\cline{2-4}
&Brushless DC motors \& controllers&7/8&$\sbullet$ For frequent use on long lifetime missions\\\hline
Thermal control&Active or passive&8&$\sbullet$ Flight heritage on robotic/human missions\\\hline
Solar sails & Current sail materials &8&
$\sbullet$ Several sailcraft flown in Earth's orbit and near Venus\\
&Advanced sail materials&3/4&$\sbullet$ No sails yet flown at 0.2 AU from the Sun; needs TDM\\\hline
Electrical power:&Solar panel arrays&7/8&$\sbullet$~For medium distance from the Sun\\
power generation&&&$\sbullet$~May exploit ultrathin solar cells on sail\\\cline{2-4}
&Miniature radioisotope power units&4/7&$\sbullet$ For large heliocentric distances\\[-3pt]
&&&\phantom{$\sbullet$~}(miniaturized/currently available)\\\hline
Electrical power:&High specific energy batteries&7&$\sbullet$ Secondary battery\\
power storage&&&$\sbullet$ High energy, low mass\\\cline{2-4}
&Supercapacitors&4/5&$\sbullet$~Very long-term storage\\\hline
Attitude determination&Star trackers, inertial measurement&7/8&$\sbullet$ Autonomy during perihelion\\
&units and Sun sensors&&$\sbullet$ Need thermal protection near the Sun\\\hline
Attitude control&Reaction \& momentum wheels,&7/8&$\sbullet$ Lightweight wheels suited for small spacecraft\\
&control moment gyroscopes&&$\sbullet$ Gyros suited for large spacecraft,\\[-3pt]
&&&\phantom{$\sbullet$~}heavy but power efficient\\\cline{2-4}
&Vanes (solar sails)&4/6&$\sbullet$ Performance for attitude control \\[-3pt]
&&&\phantom{$\sbullet$~}yet to be demonstrated\\
&&&$\sbullet$ JWST deployed 5 layers (772~m$^2$) of thin-film material\\\hline
In-flight aggregation&
Precision nav/attitude, docking&6/7&$\sbullet$ Docking for microsats (NASA CPOD mission)\\
&&&$\sbullet$ Cooperative robotic networks (NASA CARDE\footnote{Cooperative Autonomous Distributed Robotic Exploration (CADRE), see \url{https://www.nasa.gov/directorates/spacetech/game_changing_development/projects/CADRE}} mission)\\\cline{2-4}
&  Post-perihelion velocity control &3/4&$\sbullet$ Assembly of the modules with post-solar flyby velocity  \\
&   & & \phantom{~}  dispersion errors is understood; needs a flight demo \\\hline
Orbit control&Vanes (solar sails)&3/4&$\sbullet$ Single sails have flight heritage for control\\
&&&$\sbullet$ Vanes are at concept level\\\cline{2-4}
&Chemical/electrical propulsion&8&$\sbullet$ Support to vanes, mission-dependent application\\\hline
On-board computing &COTS\footnote{Commercial off-the-shelf (COTS) products} smallsat OBC&8&$\sbullet$ Deep space qualified, rad hard, COTS\\
&Modular OBC&5&$\sbullet$ Mission specific configuration\\\hline
On-board software&Spacecraft management&7&$\sbullet$ Advanced fault detection, isolation, recovery\\[-3pt]
&&&\phantom{$\sbullet$ }under development\\\cline{2-4}
&Payload data management,&3/7&$\sbullet$ Advanced data processing\\[-3pt]
&attitude \& orbit control&&\phantom{$\sbullet$~}attitude \& orbit control through vanes\\\hline
Communications&Transceivers&5/8&$\sbullet$ Miniature high-speed deep space transceivers\\[-3pt]
&&&\phantom{$\sbullet$~}under development\\\cline{2-4}
&Patch array antennas&7&$\sbullet$ For low to medium distance from Earth\\\cline{2-4}
&Reflectarray antennas&7&$\sbullet$ May be integrated with the sail\\\cline{2-4}
&Dish antennas&7&$\sbullet$ For large distances from the Earth\\\hline
\end{tabular}
\end{table*}

\section{Near-term developments}
\label{app:nearterm}

Our objective is to reach a spacecraft velocity of 5--7~AU/yr and thereby surpass the current record set by Voyager 1 (launched in 1977), traveling at a heliocentric velocity of 3.1~AU/yr. The demonstration will be a first step in revolutionizing space exploration to make it low cost, accessible and fast. In addition, the first flight will bring novel and exciting data about the nature of the Sun. The core enabling technology for setting a new velocity record is solar sail propulsion.

A small spacecraft equipped with a solar sail will get close to the Sun ($\sim$0.25~AU) and then slingshot onto a hyperbolic orbit with 5--7~AU/yr excess velocity. This sailcraft needs to be very lightweight, with a sail area to overall spacecraft area-to-mass ratio more than 45-50~m${}^2$/kg. Such a sailcraft has three key elements: sail film, support booms, and spacecraft bus with avionics and instrumentation. These elements should be carefully designed to meet a very stringent mass budget.

Preliminary studies suggest that a point design may be with a sail area of $\sim$100--144~m${}^2$ and overall spacecraft mass of $\sim$4.2--6.4~kg. This size is enabled by utilizing cubesat technology. Such a sailcraft will take $\sim$6--7~months to reach perihelion. During the entire mission, and especially during perihelion, the sail must be controlled and oriented with respect to the sun to yield an optimal trajectory and to maximize exit velocity.

In the scope of two different NASA funded programs (NIAC II \cite{Davoyan2021b}, NIAC III \cite{Turyshev-etal:2020-PhaseIII}) the design of the ``Sundiver'' sailcraft was considered at various depths of detail\footnote{In addition, relevant light sail material developments are currently conducted  at Caltech and UCLA, in part to support the Breakthrough Initiatives' StarShot program, for details see \url{https://breakthroughinitiatives.org/news/4}.}. The studies have shown the feasibility of the concept. Specifically, it was shown that sail materials surviving 0.2~AU exist, preliminary thermal modeling was conducted, and concurrent engineering trades to meet desired mass budget were performed. At the same time, detailed engineering studies are needed to determine and validate a path to building and launching, thus moving the effort from the current stage to deployment to successful mission implementation.

To enable this progress, further development steps are needed:
\begin{itemize}
\item{\bf\em Mission requirements}: Once a particular mission objective is chosen, target mission requirements (i.e., instruments, pointing precision, duration and trajectory) will guide the overall design, mission development and sailcraft building effort.
\item{\bf\em Sail materials}: Our work has shown that materials for 0.2~AU can be manufactured. These materials weigh $\sim$1.5~g/m${}^2$ (150~g for a 100~m${}^2$ sail), and have properties close to those used in current sails. Further steps require scaling of the materials to large areas (assessment shows that roll-to-roll processes can be used for this purpose). In addition, testing of large area materials for stowage and deployment is needed to ensure integrity.
\item{\bf\em Sailcraft controls}: Preliminary analysis has considered several ADCS approaches, including reaction wheels, articulable vanes, reflectivity control devices. To ensure simplicity of the first mission, a set of reaction wheels (each $\sim$120~g) is likely to be employed (these were successfully demonstrated to work with the Lightsail 2 to change its orientation on orbit). Although initial assessment was performed, a more detailed modeling connecting trajectories, sail excitation/bending modes, and ADCS is needed (such modeling is available at JPL and MSFC and is being developed at UCLA).
\item{\bf\em Sailcraft booms and deployment}: For such a relatively small area sail lightweight composite booms can be utilized ($<20$~g/m), 560--670~g total. Designs that minimize deployment mass (i.e., mass of the deployment mechanisms). Furthermore, detailed trades between boom mass and sail control authority (sail rigidity) should be performed. Boom manufacturing at scale needed ($\sim$7~m long) and integration with sail materials should be assessed. Lab scale stowage and deployment tests should be performed.
\item{\bf\em Sailcraft bus}: The limited mass budget calls for a careful design and choice of the bus and other spacecraft systems. If $\sim$1.1~kg is allocated for sail materials, controls, and booms, for a 100~m${}^2$ sail the bus budget should not exceed 700~g. Careful mass allocation (and related integration) for power and communications systems is required, in addition to simple instrumentation (cameras, magnetometers). A detailed engineering model (coupled with mechanical and thermal modeling) should be performed. In addition,  engineering model must be built and tested in the relevant environment.
\end{itemize}


\end{document}